\documentclass[sort&compress]{aip-cp}

\usepackage[numbers]{natbib}
\usepackage{rotating}
\usepackage{graphicx}
\usepackage{color}

\usepackage{amssymb}
\usepackage{subeqn}
\usepackage{bm}

\def\bal#1\eal{\begin{align}#1\end{align}}
\newcommand\beq{\begin{equation}}
\newcommand\eeq{\end{equation}}
\newcommand\beqa{\begin{eqnarray}}
\newcommand\eeqa{\end{eqnarray}}

\newcommand{\text}[1]{\textrm{\scriptsize{#1}}}
\newcommand{\eqref}[1]{(\ref{#1})}
\newcommand{\vvss}{\vspace{-.5mm}\\}

\newcommand{\ed}{\end{document}}
\newcommand{\ma}{m_i}
\newcommand{\mb}{m_j}
\newcommand{\cca}{\mathbf{v}_i}
\newcommand{\ca}{\varv_i}
\newcommand{\ccb}{\mathbf{v}_j}
\newcommand{\cb}{\varv_j}

\newcommand{\JJ}{\mathbf{Q}_{ij}}

\newcommand{\kk}{\widehat{\bm{\sigma}}}
\newcommand{\wwa}{\bm{\omega}_i}
\newcommand{\wwb}{\bm{\omega}_j}
\newcommand{\wa}{{\omega}_i}
\newcommand{\wb}{{\omega}_j}

\newcommand{\Ia}{I_i}
\newcommand{\Ib}{I_j}
\newcommand{\da}{\sigma_i}
\newcommand{\db}{\sigma_j}

\newcommand{\dab}{\sigma_{ij}}
\newcommand{\x}{\times}
\newcommand{\gh}{\mathbf{v}_{ij}}
\newcommand{\g}{\varv_{ij}}
\newcommand{\een}{\alpha_{ij}}

\newcommand{\eet}{\beta_{ij}}

\newcommand{\enn}{\overline{\alpha}_{ij}}

\newcommand{\ett}{\overline{\beta}_{ij}}
\newcommand{\mab}{m_{ij}}
\newcommand{\qab}{\kappa_{ij}}
\newcommand{\qa}{\kappa_{i}}
\newcommand{\qb}{\kappa_{j}}

\newcommand{\fa}{f_{i}}
\newcommand{\fb}{f_{j}}

\newcommand{\fat}{f_{i}^\text{tr}}

\newcommand{\far}{f_{i}^\text{rot}}

\newcommand{\Tat}{T_{i}^\text{tr}}
\newcommand{\Tbt}{T_{j}^\text{tr}}
\newcommand{\Tt}{T^\text{tr}}
\newcommand{\Tar}{T_{i}^\text{rot}}
\newcommand{\Tbr}{T_{j}^\text{rot}}
\newcommand{\Tr}{T^\text{rot}}
\newcommand{\Qab}{J_{ij}}
\newcommand{\Iab}{\mathcal{J}_{ij}}

\newcommand{\na}{n_i}
\newcommand{\nb}{n_j}

\newcommand{\zabt}{\xi_{ij}^\text{tr}}
\newcommand{\zt}{\xi^\text{tr}}
\newcommand{\zabr}{\xi_{ij}^\text{rot}}
\newcommand{\zr}{\xi^\text{rot}}
\newcommand{\al}{i}
\newcommand{\be}{j}
\newcommand{\tr}{\text{tr}}
\newcommand{\rot}{\text{rot}}

\newcommand{\SSab}{\mathbf{S}_{ij}}
\newcommand{\Sab}{{S}_{ij}}
\newcommand{\chiab}{{\chi}_{ij}}

\newcommand{\llangle}{\langle\!\langle}
\newcommand{\rrangle}{\rangle\!\rangle}
\newcommand{\xx}{\mathbf{c}}
\newcommand{\dd}{\mathrm{d}}

\newcommand{\BB}{\mathfrak{B}_{ij,\kk}}
\newcommand{\dt}{d_{\text{tr}}}
\newcommand{\dr}{d_{\text{rot}}}

\begin{document}

% Title portion
\title{Energy Production Rates of Multicomponent Granular Gases of Rough Particles. A Unified View of Hard-Disk and Hard-Sphere Systems}

\author[aff1]{Alberto Meg\'ias%\noteref{note1,note2}
}
\author[aff1,aff2]{Andr\'es Santos\corref{cor1}%\noteref{note2}
}
\eaddress[url]{http://www.eweb.unex.es/eweb/fisteor/andres/Cvitae/}
%\eaddress{anotherauthor@thisaddress.yyy}

\affil[aff1]{Departamento de F\'isica, Universidad de Extremadura, 06006 Badajoz, Spain.}
\affil[aff2]{Instituto de Computaci\'on Cient\'ifica Avanzada (ICCAEx), Universidad de Extremadura, 06006 Badajoz, Spain.}
%\affil[aff3]{You would list an author's second affiliation here.}
\corresp[cor1]{Corresponding author: andres@unex.es}
%\authornote[note1]{This is an example of first authornote.}
%\authornote[note2]{This is an example of second authornote.}

\maketitle

\begin{abstract}
Granular gas mixtures modeled as systems of inelastic and rough particles, either hard disks on a plane or hard spheres, are considered. Both classes of systems are embedded in a three-dimensional space ($d=3$) but, while in the hard-sphere case the translational and angular velocities are vectors with the same dimensionality (and thus there are $\dt=3$ translational and $\dr=3$ rotational degrees of freedom), in the hard-disk case the translational velocity vectors are planar (i.e., $\dt=2$ translational degrees of freedom) and the angular velocity vectors are orthogonal to the motion plane (i.e., $\dr=1$ rotational degree of freedom). This complicates a unified presentation of both classes of systems, in contrast to what happens for smooth, spinless particles, where a treatment of $d$-dimensional spheres is possible.
In this paper, a kinetic-theory derivation of the (collisional) energy  production rates  $\xi_{ij}^{\text{tr}}$ and  $\xi_{ij}^{\text{rot}}$ (where the indices $i$ and $j$ label different components)  in terms of the numbers of degrees of freedom $\dt$ and $\dr$ is presented. Known hard-sphere and hard-disk expressions are recovered by particularizing to $(\dt,\dr)=(3,3)$ and $(\dt,\dr)=(2,1)$, respectively. Moreover, in the case of spinless particles with $d=\dt$, known energy production rates  $\xi_{ij}^{\text{tr}}=\xi_{ij}$ of smooth $d$-dimensional spheres are also recovered.
\end{abstract}

% Head 1
\section{INTRODUCTION}

A ``gas'' made of identical and smooth hard disks or spheres with a constant coefficient of normal restitution is perhaps the simplest and most widely used model of a granular gas \cite{D00,OK00,PL01,G03,K04,BP04,AT06,RN08}. On the other hand, the mesoscopic or macroscopic nature of the ``grains'' may ask for a refinement of the model by allowing for particle-particle surface friction or ``roughness'' (usually accounted for by a constant coefficient of tangential restitution)
\cite{JR85a,LS87,C89,L91,LB94,GS95,L95,L96,ZTPSH98,HZ97,ML98,LHMZ98,HHZ00,AHZ01,MHN02,CLH02,JZ02,PZMZ02,MSS04,HCZHL05,GNB05,Z06,BPKZ07,GA08,KBPZ09,K10a%
,S11a,SKS11,SK12,MDHEH13,VSK14,VSK14b,KSG14,RA14,VS15,FH17,SP17,DF17,GSK18},
polydispersity (i.e., assuming that the particles belong to more than one component, each one characterized by different mechanical properties)
\cite{JM89,GD99b,HQL01,JY02,MG02b,BT02a,DHGD02,GD02,G02,BRM05,SA05,SGNT06,GDH07,GHD07,GM07,VGS07,G08,G08b,UKAZ09}, or both \cite{VT04,PTV07,CP08,SKG10,S11b,VLSG17,VLSG17b,S18}.

An interesting feature of multicomponent gases of rough disks or spheres is the general breakdown of energy equipartition, even in homogeneous and isotropic states (driven or undriven). This is characterized by unequal translational ($\Tat$) and rotational ($\Tar$) temperatures associated with each component $i$. The rate of change of the  translational (rotational) mean kinetic energy of particles of component $i$ due to collisions with particles of component $j$ defines the energy production rate $\zabt$ ($\zabr$). By means of
kinetic-theory tools, the production rates $\zabt$ and $\zabr$ have been derived separately for disks \cite{S18} and spheres \cite{SKG10}  as functions of $\Tat$, $\Tbt$, $\Tar$, $\Tbr$, and of the mechanical parameters (masses, diameters, moments of inertia, and coefficients of normal and tangential restitution) for each pair $ij$.

Whereas in the case of smooth, spinless particles a generic kinetic-theory treatment of $d$-dimensional hard spheres is possible \cite{BC01,G02,SA05,GM07,VGS07}, this is far less straightforward if particles have a rotational or angular motion, in addition to the translational motion of the center of mass. In fact, mathematical operations such as the cross product of \emph{two} vectors (and hence mechanical quantities such as angular momentum and torque) are, in general, meaningful in a three-dimensional space ($d=3$) only. Furthermore, the existence of surface friction or roughness establishes a neat separation between the cases of disks on a plane and spheres. Both classes of particles are embedded in a common three-dimensional space, but spinning spheres have $\dt=3$ translational plus $\dr=3$ rotational degrees of freedom, while spinning disks on a plane have $\dt=2$ translational and $\dr=1$ rotational degrees of freedom.

The aim of this work is to unify the derivations of $\zabt$ and $\zabr$ for disks \cite{S18} and spheres \cite{SKG10} so that they depend parametrically on both $\dt$ and $\dr$. On the one hand, particularization to smooth particles with $\dt=d$ allows us to recover known results for an arbitrary number $d$ of spatial dimensions \cite{G02,SA05,GM07,VGS07}. On the other hand, in the case of rough particles, particularization to $(\dt,\dr)=(2,1)$ and $(\dt,\dr)=(3,3)$ recovers previous results for disks \cite{S18} and spheres \cite{SKG10}, respectively.

\section{BINARY COLLISIONS}\label{BC}
\begin{figure}[h]
\centerline{\includegraphics[scale=0.4]{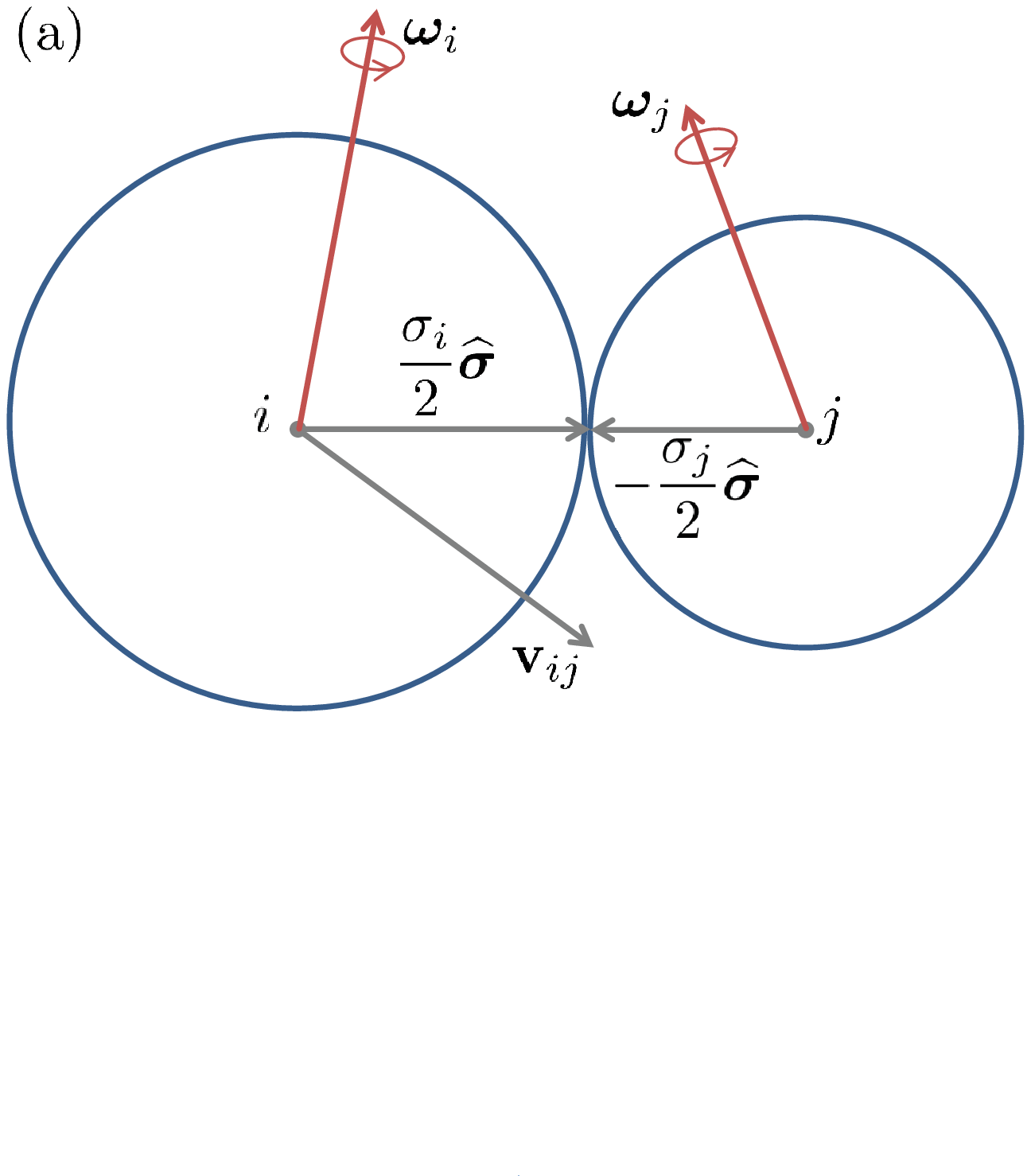}\hspace{2cm}\includegraphics[scale=0.4]{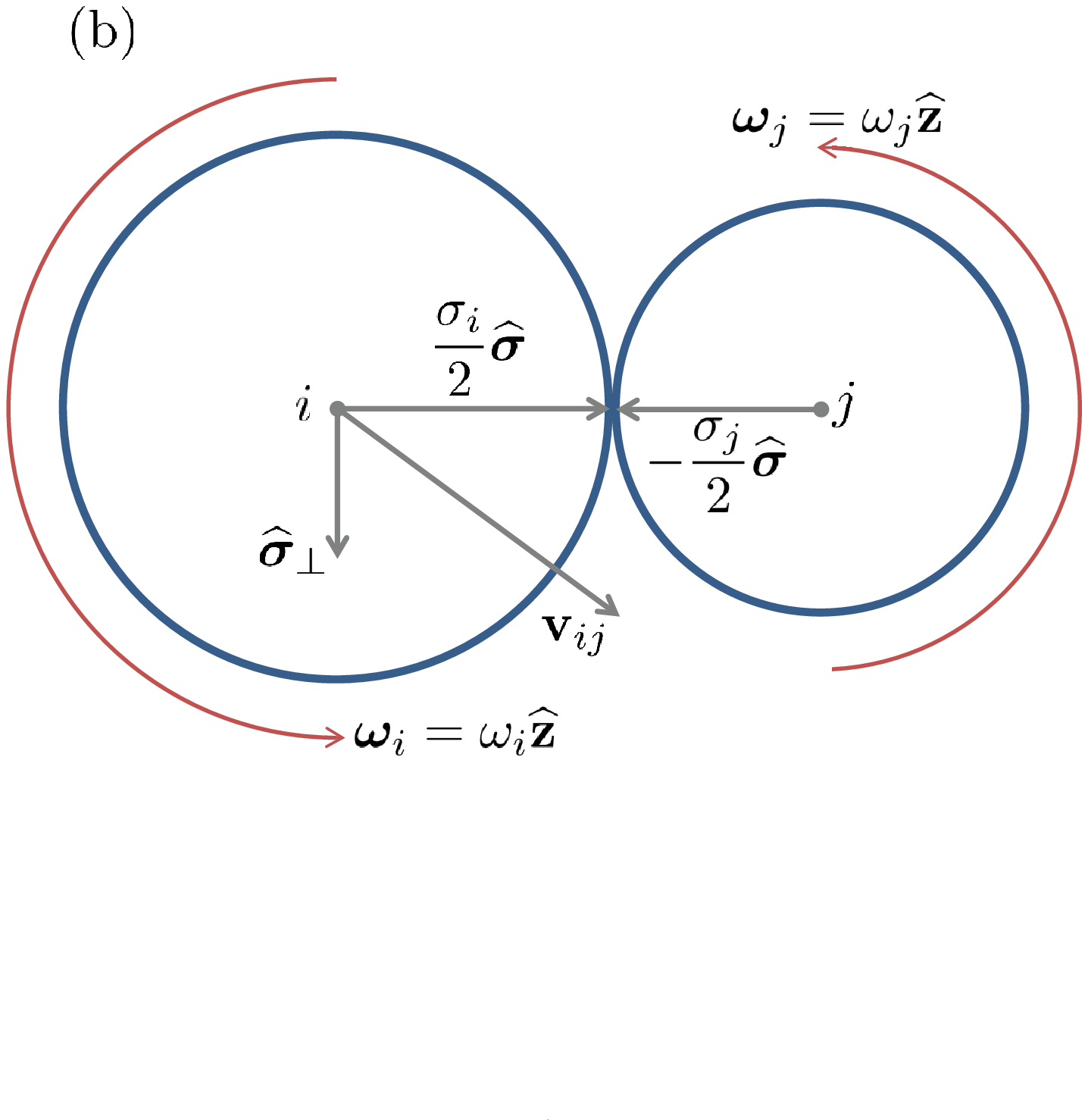}}
\caption{Sketch of the precollisional quantities of particles $i$ and $j$ in the frame of reference solidary with particle $j$. In panel (a)
(hard spheres), the angular velocities $\wwa$ and $\wwb$ can point in any direction of the three-dimensional space. In panel (b) (hard disks), the angular velocities $\wwa$ and $\wwb$ are orthogonal to the plane ($xy$) of translational motion.}
\label{sketch}
\end{figure}

Let us consider the binary collision of two hard spheres (or disks)  of masses $\ma$ and $\mb$, diameters $\da$ and $\db$, and moments of inertia $\Ia$ and $\Ib$. Before collision, the particles rotate with angular velocities $\wwa$ and $\wwb$, while their respective centers of mass move with translational velocities $\cca$ and $\ccb$. This is sketched in Fig.\ \ref{sketch}, where
$\widehat{\boldsymbol{\sigma}}\equiv(\mathbf{r}_j-\mathbf{r}_i)/|\mathbf{r}_j-\mathbf{r}_i|$ is a unit vector pointing from the center of sphere $i$ to the center of sphere $j$ and  $\mathbf{v}_{ij}=\mathbf{v}_i-\mathbf{v}_j$ is the relative velocity of the centers of mass.
The relative velocity ($\mathbf{w}_{ij}$) of the points of the spheres or disks which are in contact at the collision is
\beq
\mathbf{w}_{ij}=\mathbf{v}_{ij}-\kk\x\SSab, \quad \SSab\equiv \frac{\da}{2}\wwa+\frac{\db}{2}\wwb.
\label{wvel_rel}
\eeq
In the case of spheres, the vector $\kk\x\SSab$  points in any direction of the three-dimensional space. In the case of disks, however, $\kk\x\SSab=\Sab\kk_\perp$ lies on the plane of motion, where $\kk_\perp=\kk\times\widehat{\mathbf{z}}=\widehat{\sigma}_y\widehat{\mathbf{x}}-\widehat{\sigma}_x\widehat{\mathbf{y}}$.

The relative velocity $\mathbf{w}_{ij}$ can be decomposed into a normal component (parallel to $\kk$) and a tangential component  (orthogonal to $\kk$):
\beq
\mathbf{w}_{ij}=(\mathbf{w}_{ij}\cdot\kk)\kk-\kk\x(\kk\x\mathbf{w}_{ij}).
\eeq
In the case of disks, $-\kk\x(\kk\x\mathbf{w}_{ij})=(\mathbf{w}_{ij}\cdot\kk_\perp)\kk_\perp$. After collision, the normal and tangential components of $\mathbf{w}_{ij}$ are modified by constant factors $\een$ (coefficient of normal restitution) and $\eet$ (coefficient of tangential restitution), respectively, i.e.,
\begin{equation}
\BB\mathbf{w}_{ij}\cdot\widehat{\boldsymbol{\sigma}}=-\een\mathbf{w}_{ij}\cdot\widehat{\boldsymbol{\sigma}},\quad \BB\kk\x\mathbf{w}_{ij}=-\eet\kk\x\mathbf{w}_{ij},
\label{walpha}
\end{equation}
where the operator $\BB$ acting on a precollisional quantity gives the associated postcollisional quantity as the result of a collision with unit vector $\kk$ between particles of components $i$ and $j$.  The coefficient of normal restitution ranges from $\een=0$ (perfectly inelastic particles) to $\een=1$ (perfectly elastic particles), while the coefficient of tangential restitution ranges from $\eet=-1$ (perfectly smooth particles) to  $\eet=1$ (perfectly rough particles).

Equation \eqref{walpha}, together with the laws of conservation of linear and angular momenta, yield the following collision rules \cite{SKG10,S18,Z06}
\beq
\BB\mathbf{v}_i=\mathbf{v}_i-\frac{1}{\ma}\mathbf{Q}_{ij},\quad \BB\mathbf{v}_j=\mathbf{v}_j+\frac{1}{\mb}\mathbf{Q}_{ij},
\quad \BB\wwa=\wwa-\frac{\da}{2\Ia}\kk\x\mathbf{Q}_{ij},\quad \BB\wwb=\wwb-\frac{\db}{2\Ib}\kk\x\mathbf{Q}_{ij},
\label{cons_lin_vs}
\eeq
where
\begin{equation}
\JJ=\mab\enn(\gh\cdot\kk)\kk+\mab\ett\left[\gh-(\gh\cdot\kk)\kk-\kk\x\SSab\right].
\label{JJ}
\end{equation}
Here,
\begin{equation}
\mab\equiv \frac{\ma\mb}{\ma+\mb},\quad\enn\equiv1+\een,\quad\ett\equiv\frac{\qab}{1+\qab}\left(1+\eet\right),\quad \qab\equiv \qa\qb\frac{\ma+\mb}{\qa\ma+\qb\mb},\quad \qa\equiv\frac{4\Ia}{\ma\da^2}, \quad \qb\equiv\frac{4\Ib}{\mb\db^2}.
\label{20}
\end{equation}
The values of the reduced moments of inertia $\kappa_i$ run from $\kappa_i=0$, if the mass is completely concentrated in the center of the body, to $\kappa_i=1$ (disks) or $\kappa_i=\frac{2}{3}$ (spheres), if the mass is concentrated on the perimeter of the particle. In the case of a uniform mass distribution,  $\kappa_i=\frac{1}{2}$  (disks) or $\kappa_i=\frac{2}{5}$ (spheres).
Note that for perfectly smooth particles ($\eet=-1$) one has $\ett=0$, and thus the angular velocities are not affected by collisions.

The rules for restituting collisions are
\beq
\BB^{-1}\mathbf{v}_i=\mathbf{v}_i-\frac{1}{\ma}\mathbf{Q}_{ij}^{-}, \quad \BB^{-1}\mathbf{v}_j=\mathbf{v}_j+\frac{1}{\mb}\mathbf{Q}_{ij}^{-},
\quad \BB^{-1}\wwa=\wwa-\frac{\da}{2\Ia}\kk\x\mathbf{Q}_{ij}^{-},\quad \BB^{-1}\wwb=\wwb-\frac{\db}{2\Ib}\kk\x\mathbf{Q}_{ij}^{-},
\label{resw}
\eeq
where
\begin{equation}
\mathbf{Q}_{ij}^{-}=\mab\frac{\enn}{\een}(\gh\cdot\kk)\kk+\mab\frac{\ett}{\eet}\left[\gh-(\gh\cdot\kk)\kk -\kk\x\SSab\right].
\label{Qij-}
\end{equation}

\begin{table}[t]
\caption{Relevant collisional changes.}
\label{table:0}
\tabcolsep7pt\begin{tabular}{rcl}
\hline
$\psi_{ij}(\xx_i,\xx_j)$ && \hspace{4cm} $\left(\BB-1\right)\psi_{ij}(\xx_i,\xx_j)$\\
\hline
\vvss
$\ma\cca$&&$\displaystyle{-\mab\enn(\gh\cdot\kk)\kk-\mab\ett\left[\gh-(\gh\cdot\kk)\kk -\kk\x\SSab\right]}$\\
\vvss
$\Ia\wwa$&&$\displaystyle{-\frac{\mab\da}{2}\ett\left[\kk\x\gh+\SSab-(\SSab\cdot\kk)\kk\right]}$\\
\vvss
$\ma\ca^2$&&$\displaystyle{\frac{\mab^2\enn^2}{\ma}(\gh\cdot\kk)^2-2\mab\enn(\gh\cdot\kk)(\cca\cdot\kk)
+\frac{\mab^2\ett^2}{\ma}\left[(\kk\x\gh)^2+(\kk\x\SSab)^2+2(\kk\x\gh)\cdot\SSab\right]
}$\\
&&\hspace{0.5cm}$\displaystyle{-2\mab\ett\left[(\kk\x\gh)\cdot(\kk\x\cca)+(\kk\x\cca)\cdot\SSab\right]}$\\
\vvss
$\Ia\wa^2$&&$\displaystyle{\frac{\mab^2\ett^2}{\ma\qa}\left[(\kk\x\gh)^2+(\kk\x\SSab)^2
+2(\kk\x\gh)\cdot\SSab\right] -\mab\ett\da\left[(\kk\x\SSab)\cdot(\kk\x\wwa)+(\kk\x\gh)\cdot\wwa\right]
}$\\
\vvss
$\ma\ca^2+\mb\cb^2$&&$\displaystyle{-\mab\left(1-\een^2\right)(\gh\cdot\kk)^2
+\mab\ett^2\left[(\kk\x\gh)^2+(\kk\x\SSab)^2+2(\kk\x\gh)\cdot\SSab\right]
}$\\
&&\hspace{0.5cm}$\displaystyle{-2\mab\ett\left[(\kk\x\gh)^2+(\kk\x\gh)\cdot\SSab\right]}$\\
\vvss
$\Ia\wa^2+\Ib\wb^2$&&$\displaystyle{\frac{\mab\ett^2}{\qab}\left[(\kk\x\gh)^2+(\kk\x\SSab)^2
+2(\kk\x\gh)\cdot\SSab\right] -2\mab\ett\left[(\kk\x\SSab)^2+(\kk\x\gh)\cdot\SSab\right]}$\\
\vvss
$E_{ij}$&&$\displaystyle{-{\mab}\frac{1-\een^2}{2}(\gh\cdot\kk)^2-\mab\frac{\qab}{1+\qab}\frac{1-\eet^2}{2}\left[(\kk\x\gh)^2
+(\kk\x\SSab)^2+2(\kk\x\gh)\cdot\SSab \right] }$\\
\hline
\end{tabular}
\end{table}

Given a certain dynamic variable $\psi_{ij}(\xx_i,\xx_j)$, where the short-hand notation $\xx_i\equiv\{\cca,\wwa\}$, $\xx_j\equiv\{\ccb,\wwb\}$ has been introduced, Eqs.\ \eqref{cons_lin_vs} and \eqref{JJ} provide its collisional change $\left(\BB-1\right)\psi_{ij}(\xx_i,\xx_j)$. The most relevant cases are presented in Table \ref{table:0} \cite{SKG10}. In the last row,
\begin{equation}
E_{ij}=\frac{1}{2}\ma\ca^2+\frac{1}{2}\mb\cb^2+\frac{1}{2}\Ia\wa^2+\frac{1}{2}\Ib\wb^2
\label{Eij}
\end{equation}
is the total kinetic energy (translational plus rotational) of both colliding particles. Energy is conserved only if the particles are elastic  ($\een=1$)  and  either perfectly smooth ($\eet=-1$) or perfectly rough ($\eet=1$). Otherwise, $\left(\BB-1\right)E_{ij}<0$ and kinetic energy is dissipated upon collisions.

The collision rules \eqref{cons_lin_vs} and \eqref{JJ}, as well as those in Table \ref{table:0}, hold both for spheres and disks. In the latter case, however,  some terms may  simplify \cite{S18}. On the other hand, the Jacobian of the transformation $\{\xx_i,\xx_j\}\to \{\BB\xx_i,\BB\xx_j\}$ is different for spheres ($\dt=\dr=3$) and disks ($\dt=2$, $\dr=1$), namely \cite{SKG10,S18}
\begin{equation}
\mathfrak{J}_{ij}\equiv \left|\frac{\partial\left(\BB\xx_i,\BB\xx_j\right)}{\partial\left(\xx_i,\xx_j\right)}\right|=
\left|\frac{\partial\left(\xx_i,\xx_j\right)}{\partial\left(\BB^{-1}\xx_i,\BB^{-1}\xx_j\right)}\right|=
\left\{
\begin{array}{ll}
\een|\eet|&\mathrm{(disks)},\\
\een\eet^2&\mathrm{(spheres)}.
\end{array}
\right.
\label{Jac3D}
\end{equation}

\section{BOLTZMANN EQUATION}
Let $f_i(\mathbf{r},\xx_i;t)$ be the one-body velocity distribution function of particles of component $i$. In the low-density limit, by application of the molecular chaos assumption on the first equation of the Bogoliubov--Born--Green--Kirkwood--Yvon (BBGKY) hierarchy,  $f_i(\mathbf{r},\xx_i;t)$ obeys the Boltzmann equation \cite{SKG10,S18}
\begin{equation}
\partial_t \fa(\mathbf{r},\xx_i;t)+\cca\cdot\nabla \fa (\mathbf{r},\xx_i;t)
=\sum_{\be} \Qab[\mathbf{r},\xx_i;t|\fa,\fb],
\label{2}
\end{equation}
where
\beq
\label{Jij3D}
\Qab[\mathbf{r},\xx_i;t|\fa,\fb]=\chi_{ij}\dab^{\dt-1}\int \dd\xx_j\int_{+}\dd\kk\, (\gh\cdot\kk)\left(\frac{1}{\een\mathfrak{J}_{ij}} \BB^{-1}-1\right)\fa(\mathbf{r},\mathbf{c}_i;t)\fb(\mathbf{r},\mathbf{c}_j;t)
\eeq
is the bilinear Boltzmann collision operator. In Eq.\ \eqref{Jij3D}, $\chiab$ is the contact value of the pair correlation function,  $\dab\equiv \frac{1}{2}(\da+\db)$, $\int \dd\xx_j\equiv\int\dd\ccb\int\dd\wwb$, and  $\int_+ d\kk\,\equiv \int d\kk\,\Theta(\gh\cdot\kk)$, $\Theta(x)$ being the Heaviside step function. The following mathematical integrals over $\kk$ will be needed  \cite{vNE98}:
\begin{subequations}
\label{An1}
\begin{equation}
\label{An1a}
\int_{+} \dd\kk \,(\gh\cdot\kk)^{\ell}=B_\ell\g^\ell,\quad
\int_{+} \dd\kk \,(\gh\cdot\kk)^\ell\kk=B_{\ell+1}\g^{\ell-1}\gh,
\end{equation}
\begin{equation}
\label{An1b}
\quad \int_{+} \dd\kk \,(\gh\cdot\kk)^\ell\kk\kk=B_{\ell+2}\g^{\ell-2}\gh\gh+\frac{B_\ell-B_{\ell+2}}{\dt-1}\g^{\ell-2}\left(\g^2\mathsf{I}_{\tr}-\gh\gh \right),
\end{equation}
\end{subequations}
where  $\mathsf{I}_{\tr}$ is the $\dt \x \dt$ unit tensor and $B_\ell=\pi^{(\dt-1)/2}\Gamma(\ell/2+1/2)/\Gamma(\ell/2+\dt/2)$.

Given a one-body dynamic variable $\psi_i(\xx_i)$, its average value is
\begin{equation}
\langle \psi_i(\xx_i)\rangle\equiv \frac{1}{\na}\int \dd\xx_i\, \psi_i(\xx_i) \fa(\xx_i), \quad \na=\int \dd\xx_i\,  \fa(\xx_i),\quad n=\sum_i n_i,
\label{III.3}
\end{equation}
$\na$ and $n$
being the number density of component $\al$ and the total number density, respectively. For the sake of brevity, in Eq.\ \eqref{III.3} and henceforth  the spatial and temporal arguments are omitted.
Analogously, the average of a two-body dynamic variable $\psi_{ij}(\xx_i,\xx_j)$ is
\begin{equation}
\llangle \psi_{i,j}(\xx_i,\xx_j)\rrangle\equiv \frac{1}{\na \nb}\int \dd\xx_i \int \dd\xx_j\,\psi_{ij}(\xx_i,\xx_j)\fa(\xx_i)\fb(\xx_j).
\label{17}
\end{equation}
Multiplying both sides of the Boltzmann equation \eqref{2} by $\psi_i(\xx_i)$ and integrating over $\xx_i$, we obtain the balance equation
\begin{equation}
\frac{\partial}{\partial t}\left[n_i\langle \psi_i(\xx_i) \rangle\right]+\nabla\cdot\left[n_i\langle \mathbf{v}_i\psi_i(\xx_i)\rangle\right]=\sum_j\mathcal{J}_{ij}[\psi_i|f_i,f_j],
\label{balance}
\end{equation}
where
\beq
\label{Jcalij}
\Iab[\psi_i|\fa,\fb]\equiv\int \dd\xx_i\, \psi_i(\xx_i)J_{ij}[\xx_i|\fa,\fb]=\chi_{ij}\dab^{\dt-1}\int \dd\xx_i \int \dd\xx_j \int_{+} \dd\kk \,(\gh\cdot\kk)\fa(\xx_i)\fb(\xx_j)\left(\BB-1\right)\psi_i(\xx_i).
\eeq
Therefore, ${n_i}^{-1}\Iab[\psi_i|\fa,\fb]=\left.\partial_t\langle\psi_i\rangle\right|_{\text{coll},j}$ represents the rate of change of the quantity $\psi_i(\xx_i)$ due to collisions with particles of component $j$. Analogously, in the case of a two-body dynamic variable $\psi_{ij}(\xx_i,\xx_j)$, the collisional production rate $\Iab[\psi_{ij}|\fa,\fb]$ is defined by the second equality of Eq.\ \eqref{Jcalij} with the replacement $\psi_i(\xx_i)\to \psi_{ij}(\xx_i,\xx_j)$.

\begin{table}[t]
\caption{Relevant collisional integrals  in terms of two-body averages.}
\label{table:1}
\tabcolsep7pt\begin{tabular}{rcl}
\hline
$\psi_{ij}(\xx_i,\xx_j)$ && \hspace{4cm} $\mathcal{K}_{ij}[\psi_{ij}|\fa,\fb]$\\
\hline
\vvss
$\ma\cca$&&$\displaystyle{\left(\enn+\frac{\dt-1}{2}\ett\right)\llangle \g\gh\rrangle-\frac{\sqrt{\pi}\Gamma(3/2+\dt/2)}{2\Gamma(1+\dt/2)} \ett\llangle\gh \times \SSab\rrangle}$\\
\vvss
$\Ia\wwa$&&$\displaystyle{\frac{\da}{4}\ett  \left[3\llangle \g\SSab\rrangle-  \llangle \g^{-1}(\gh\cdot\SSab)\gh \rrangle \right]}$\\
\vvss
$\ma\ca^2$&&$\displaystyle{2\left(\enn+\frac{\dt-1}{2}\ett\right)\llangle \g(\cca\cdot\gh)\rrangle
-\frac{m_{ij}}{m_i}\left(\enn^2+\frac{\dt-1}{2}\ett^2\right)\llangle \g^3\rrangle
}$\\
&&\hspace{0.5cm}$\displaystyle{- \frac{\sqrt{\pi}\Gamma(3/2+\dt/2)}{\Gamma(1+\dt/2)}\ett\llangle\SSab\cdot(\cca\times\gh)\rrangle
-{\frac{\mab\ett^2}{2\ma}\left[3\llangle \g\Sab^2\rrangle- \llangle \g^{-1}(\gh\cdot\SSab)^2\rrangle\right]}}$\\
\vvss
$\Ia\wa^2$&&$\displaystyle{ \frac{\da}{2}\ett\left[3\llangle\g \wwa\cdot\SSab\rrangle- \llangle \g^{-1} (\gh\cdot\SSab)(\gh\cdot\wwa)\rrangle  \right]
-\frac{\ett^2 \mab}{2\ma\qa}\left[(\dt-1)\llangle \g^3\rrangle
\right.
}$\\
&&\hspace{0.5cm}$\displaystyle{\left.
+3\llangle\g\Sab^2\rrangle -  \llangle \g^{-1}(\gh\cdot\SSab)^2\rrangle \right] }$\\
\vvss
$\ma\ca^2+\mb\cb^2$&&$\displaystyle{\left[1-\een^2+\frac{\dt-1}{2}\ett\left(2-\ett\right)\right]\llangle \g^3 \rrangle-\frac{\ett^2}{2}{\left[3\llangle \g\Sab^2\rrangle- \llangle \g^{-1}(\gh\cdot\SSab)^2\rrangle\right]}
}$\\
\vvss
$\Ia\wa^2+\Ib\wb^2$&&$\displaystyle{\frac{\ett}{2\qab}\left(2\qab-\ett\right)\left[3\llangle \g\Sab^2\rrangle- \llangle \g^{-1}(\gh\cdot\SSab)^2\rrangle \right]
- \frac{\dt-1}{2}\frac{\ett^2}{\qab}\llangle \g^3\rrangle}$\\
\vvss
$E_{ij}$&&$\displaystyle{\frac{1-\een^2}{2}\llangle \g^3\rrangle+\frac{\qab(1-\eet^2)}{4(1+\qab)}\left[(\dt-1)\llangle \g^3 \rrangle + 3\llangle \g \Sab^2\rrangle- \llangle \g^{-1} (\gh\cdot \SSab)^2 \rrangle\right]}$\\
\hline
\end{tabular}
\end{table}

Let us focus on the quantities $\psi_{ij}(\xx_i,\xx_j)$ listed on the first column of Table \ref{table:0}. When obtaining the corresponding production rates $\Iab[\psi_{ij}|\fa,\fb]$, the angular integrals $\int_{+}\dd\kk \,(\gh\cdot\kk)\left(\BB-1\right)\psi_{ij}(\xx_i,\xx_j)$ can be evaluated with the help of Eqs.\ \eqref{An1}, so that $\Iab[\psi_{ij}|\fa,\fb]$ are expressed in terms of two-body averages. Some care is needed when applying Eq.\ \eqref{An1b} and contracting the tensor $\mathsf{I}_{\tr}$ with an angular velocity vector, for instance  $\mathsf{I}_{\tr}\cdot\SSab$. In the case of spheres ($\dt=3$), one obviously have $\mathsf{I}_{\tr}\cdot\SSab=\SSab$. However, in the case of disks ($\dt=2$) the vector $\SSab$ is orthogonal to the subspace where the identity tensor $\mathsf{I}_{\tr}$ acts, and thus $\mathsf{I}_{\tr}\cdot\SSab=0$. To unify both possibilities, and taking into account that the rotational degrees of freedom are meaningless except in the cases of disks and spheres, it is convenient to write $\mathsf{I}_{\tr}\cdot\SSab=(\dt-2)\SSab$. Analogously, $\mathsf{I}_{\tr}:\SSab\SSab=(\dt-2)\Sab^2$ and $\mathsf{I}_{\tr}:\SSab\wwa=(\dt-2)\wwa\cdot\SSab$.
The final results are displayed in Table \ref{table:1}, where we have introduced the scaled collisional integrals
\beq
{\mathcal{K}_{ij}[\psi_{ij}|\fa,\fb]\equiv-\frac{\Gamma(3/2+\dt/2)}{\pi^{(\dt-1)/2}}}\frac{\Iab[\psi_{ij}|\fa\fb]}{\chi_{ij}\mab\na\nb\dab^{\dt-1}}.
\eeq

The most important one-body averages are
\beq
\mathbf{u}_i=\langle \cca\rangle,\quad \mathbf{u}=\frac{\sum_i \ma\na\mathbf{u}_i}{\sum_i \ma\na},\quad \bm{\Omega}_i=\langle\wwa\rangle,\quad \frac{\dt}{2}\Tat=\frac{\ma}{2}\langle (\cca-\mathbf{u})^2\rangle,\quad \frac{\dr}{2}\Tar=\frac{\Ia}{2}\langle \wa^2\rangle,
\quad T=\sum_\al\frac{\na}{n}\frac{\dt\Tat+\dr\Tar}{\dt+\dr},
\eeq
where $\mathbf{u}_i$ and $\bm{\Omega}_i$ are partial flow and angular velocities, respectively, $\mathbf{u}$ is the global flow velocity,  $\Tat$ and $\Tar$ are {partial} granular temperatures associated with the translational and rotational degrees of freedom, respectively, and $T$ is the {global} granular temperature.
The partial energy production rates associated with $\Tat$ and $\Tar$ are defined as \cite{SKG10,S18}
\beq
\label{prRates}
\zabt\equiv-\frac{1}{\dt n_i \Tat}\Iab[\ma(\cca-\mathbf{u})^2|\fa,\fb], \quad \zabr\equiv-\frac{1}{\dr n_i \Tar}\Iab[\Ia\wa^2|\fa,\fb].
\eeq
These quantities can be expressed in terms of two-body averages with the help of Table \ref{table:1}. From $\zabt$ and $\zabr$ one can obtain the total production rates $\zt_\al$ and $\zr_\al$, as well as the global cooling rate $\zeta$, by the relations
\beq
\label{prRatesTotal}
\zt_{\al}\equiv-\frac{\left.\partial_t \Tat\right|_{\text{coll}}}{\Tat}=\sum_j \zabt,\quad \zr_{\al}\equiv-\frac{\left.\partial_t \Tar\right|_{\text{coll}}}{\Tar}=\sum_j \zabr,
\quad \zeta\equiv-\frac{\left.\partial_t T\right|_{\text{coll}}}{T}=\sum_i\frac{n_i\left(\dt \Tat \zt_{\al}+\dr\Tar\zr_{\al} \right)}{(\dt+\dr)nT}.
\eeq

Before closing this section, let us consider the mean collision frequency of a particle of component $i$ with particles of component $j$ as given by \cite{GS03}
\beq
\label{coll_freq}
\nu_{ij}=
\frac{1}{n_i}\chi_{ij}\dab^{\dt-1}\int \dd \xx_i\int \dd\xx_j\int_+ \dd\kk\,(\gh\cdot\kk)\fa(\xx_i)\fb (\xx_j)=\frac{\llangle \g\rrangle}{\sqrt{2}\lambda_{ij}},\quad \lambda_{ij}=\frac{\Gamma(1/2+\dt/2)}{\sqrt{2}\pi^{(\dt-1)/2}\chi_{ij}n_j\sigma_{ij}^{\dt-1}},
\eeq
where in the second equality, namely $\nu_{ij}=\llangle \g\rrangle/\sqrt{2}\lambda_{ij}$, use has been made of Eq.\ \eqref{An1a}, $\lambda_{ij}$ being the mean free path of a particle of component $i$ with respect to collisions with particles of component $j$. Note that $n_i\nu_{ij}=n_j\nu_{ji}$. The total collision frequency of a particle of component $i$ is $\nu_i=\sum_j \nu_{ij}$, while the global mean collision frequency is $\nu=\sum_i n_i\nu_i/n$.

All the results in this section are exact within the framework of the Boltzmann equation. The expressions of the main collisional integrals in terms of two-body averages are useful to evaluate those integrals by computer simulations. On the other hand, exact analytic expressions are not possible unless $\fa(\xx_i)$ and  $\fb(\xx_j)$ are known. The next section provides analytic approximations for the energy production rates based on a Maxwellian approximation.

\section{ESTIMATES OF TWO-BODY AVERAGES AND APPROXIMATE ENERGY PRODUCTION RATES}
Henceforth, we particularize to mixtures without mutual diffusion (i.e., $\mathbf{u}_i=\mathbf{u}$ for all $i$) and with  isotropic distributions of translational velocities, $\cca-\mathbf{u}$, relative to the flow velocity.

In order to get practical estimates of the two-body averages appearing in Table \ref{table:1}, let us approximate the unknown velocity distribution functions $\fa(\xx_i)$ by means of two assumptions: (i) statistical independence between translational and rotational velocities, i.e., $\fa(\xx_i)={n_i}^{-1}\fat(\cca)\far(\wwa)$, where $\fat(\cca)$ and $\far(\wwa)$ are the marginal distribution functions associated with the translational and rotational degrees of freedom, respectively; (ii) Maxwellian form for $\fat(\cca)$. Therefore,
\begin{equation}
\label{Maxw}
\fa(\xx_i)\rightarrow \left(\frac{\ma}{2\pi \Tat}\right)^{\dt/2} \exp\left[-\frac{\ma(\cca-\mathbf{u})^2}{2\Tat}\right]\far(\wwa) .
\end{equation}
When Eq.\ \eqref{Maxw}, together with  the equivalent approximation for $\fb(\xx_j)$ is inserted into Eq.\ \eqref{17}, the two-body averages appearing in Table \ref{table:1} can be explicitly evaluated in terms of the material parameters ($\ma$, $\mb$, $\da$, $\db$, $\qa$, $\qb$, $\een$, and $\eet$) and of the physical quantities $\bm{\Omega}_i$, $\bm{\Omega}_j$, $\Tat$, $\Tbt$, $\Tar$, and $\Tbr$.
Analogously to what happened with Table \ref{table:1}, the evaluation of averages involving angular velocities must be done with care to treat the cases of spheres and disks in a common setting. For instance, $\llangle \g^{-1}(\gh\cdot\SSab)\gh\rrangle=\llangle \g^{-1}\gh\gh\rrangle\cdot\llangle\SSab\rrangle=\dt^{-1}\llangle \g\rrangle\mathsf{I}_{\tr}\cdot\llangle\SSab\rrangle=\dt^{-1}(\dt-2)\llangle \g\rrangle\llangle\SSab\rrangle$. Similarly,  $\llangle \g^{-1}(\gh\cdot\SSab)^2\rrangle=\dt^{-1}(\dt-2)\llangle \g\rrangle\llangle\Sab^2\rrangle$ and
$\llangle \g^{-1}(\gh\cdot\SSab)(\gh\cdot\wwa)\rrangle=\dt^{-1}(\dt-2)\llangle \g\rrangle\llangle\wwa\cdot\SSab\rrangle$.
The results for the two-body averages are summarized in Table \ref{table:2}. Note that $\llangle \Sab^2\rrangle=
\dr\left({\Tar}/{\ma\qa}+{\Tbr}/{\mb\qb}+{\da\db {\bm{\Omega}}_\al \cdot {\bm{\Omega}}_\be}/{2\dr}\right)$ is positive definite.

\begin{table}[t]
\caption{Expressions, as obtained from the approximation \protect\eqref{Maxw}, for the two-body averages appearing in Table \ref{table:1}.}
\label{table:2}
\tabcolsep7pt\begin{tabular}{rcl}
\hline
%Quantity && Expression \\ \hline
\vvss
$\llangle \g\gh\rrangle$, $\llangle \gh\times \SSab\rrangle$, $\llangle \SSab \cdot  (\cca\x \gh)\rrangle$&&$0$\\
\vvss
$\llangle\g\rrangle$&&$\displaystyle{\frac{\sqrt{2}\Gamma\left(1/2+\dt/2\right)}{\Gamma\left({\dt}/{2}\right)}\left(\frac{\Tat}{\ma}+\frac{\Tbt}{\mb}\right)^{1/2}}$\\
\vvss
$\llangle\g^3\rrangle$&&$\displaystyle{\frac{2\sqrt{2}\Gamma\left(3/2+\dt/2\right)}{\Gamma\left({\dt}/{2}\right)}
\left(\frac{\Tat}{\ma}+\frac{\Tbt}{\mb}\right)^{3/2}}$\\
\vvss
$\llangle \g\cca\cdot\gh\rrangle$&&$\displaystyle{\frac{\Tat}{\ma}\left(\frac{\Tat}{\ma}+\frac{\Tbt}{\mb}\right)^{-1}}
\llangle\g^3\rrangle$\\
\vvss
$3\llangle \g\SSab\rrangle-\llangle \g^{-1}(\gh\cdot\SSab)\gh\rrangle$&&$ \displaystyle{\frac{\dt+1}{\dt}\left(\da {\bm{\Omega}}_\al+\db {\bm{\Omega}}_\be\right)}\llangle \g\rrangle$\\
\vvss
$3\llangle\g\Sab^2\rrangle-\llangle \g^{-1}(\gh\cdot\SSab)^2\rrangle$&&$\displaystyle{2\frac{\dt+1}{\dt}\dr\left(\frac{\Tar}{\ma\qa}+\frac{\Tbr}{\mb\qb}+\frac{\da\db {\bm{\Omega}}_\al \cdot {\bm{\Omega}}_\be}{2\dr}\right)\llangle\g\rrangle}$\\
\vvss
$3\llangle \g\wwa\cdot\SSab\rrangle-\llangle \g^{-1} (\gh\cdot\SSab)(\gh\cdot\wwa)\rrangle $&&$\displaystyle{2\frac{\dt+1}{\dt}\frac{\dr}{\da}\left(\frac{2\Tar}{\ma\qa}+\frac{\da \db \bm{\Omega}_\al \cdot \bm{\Omega}_\be }{2\dr}\right)\llangle \g\rrangle}$\\
\hline
\end{tabular}
\end{table}

\begin{table}[t]
\caption{Collisional energy production rates for polydisperse systems}
\label{table:3}
\tabcolsep7pt\begin{tabular}{rcl}
\hline
%Quantity && \hspace{5cm} Expression\\
%\hline
\vvss
$\zabt$ &&$\displaystyle{\frac{\nu_{ij}}{\dt}\frac{ 2\mab^2 }{\ma \Tat }\left\{2\left(\enn+\frac{\dr}{\dt}\ett\right)\frac{\Tat}{\mab}-\left(\enn^2+\frac{\dr}{\dt}\ett^2\right)\left(\frac{\Tat}{\ma}+\frac{\Tbt}{\mb}\right)
-\frac{\dr}{\dt}\ett^2\left(\frac{\Tar}{\ma\qa}+\frac{\Tbr}{\mb\qb}+\frac{\da\db\bm{\Omega}_{\al}\cdot\bm{\Omega}_{\be}}{2\dr} \right)\right\}
}
$\\
\vvss
$\zabr$&&$\displaystyle{\frac{\nu_{ij}}{\dt}\frac{4\mab^2\ett}{\ma\qa\Tar}\left[\frac{\Tar}{\mab}+\frac{\ma\qa}{\mab}
\frac{\da\db\bm{\Omega}_{\al}\cdot\bm{\Omega}_{\be}}{4\dr}
-\frac{\ett}{2}\left(\frac{\Tat}{\ma}+\frac{\Tbt}{\mb}+\frac{\Tar}{\ma\qa}+\frac{\Tbr}{\mb\qb}+
\frac{\da\db\bm{\Omega}_{\al}\cdot\bm{\Omega}_{\be}}{2\dr}\right) \right]
}$\\
\vvss
$\zeta$&&$\displaystyle{\sum_{\al,\be}\frac{\na\nu_{\al\be}\mab}{(\dt+\dr)nT}\left[\left(1-\een^2\right)\left(\frac{\Tat}{\ma}+\frac{\Tbt}{\mb}\right)
+\frac{\dr\qab (1-\eet^2)}{\dt(1+\qab)}
\left(\frac{\Tat}{\ma}+\frac{\Tbt}{\mb}+\frac{\Tar}{\ma\qa}+\frac{\Tbr}{\mb\qb}+\frac{\da\db \bm{\Omega}_\al \cdot \bm{\Omega}_\be}{2\dr}\right)\right]}$\\
\vvss
$\zeta_{\al\be}^\tr$&&$\displaystyle{\frac{\nu_{ij}}{\dt}\frac{2\mab^2(1-\een^2)}{ \ma\Tat}\left(\frac{\Tat}{\ma}+\frac{\Tbt}{\mb}\right)}$\\
\vvss
$\zeta_{\al\be}^\rot$&&$\displaystyle{\frac{\nu_{ij}}{\dt}\frac{2\mab^2\qab^2(1-\eet^2)}{\ma\qa(1+\qab)^2\Tar}\left(\frac{\Tat}{\ma}+\frac{\Tbt}{\mb}+
\frac{\Tar}{\ma\qa}+\frac{\Tbr}{\mb\qb}+\frac{\da\db\bm{\Omega}_i\cdot\bm{\Omega}_j}{2\dr}\right)}$\\
\vvss
$\Xi_{ij}^{(1)}$&&$\displaystyle{\frac{\nu_{ij}}{\dt}\frac{4\mab^2(1+\een)}{\ma\mb\Tat}\left(\Tat-\Tbt\right)}$\\
\vvss
$\Xi_{ij}^{(2)}$&&$\displaystyle{\frac{\nu_{ij}}{\dt^2}\frac{4\dr\mab\qab(1+\eet)}{\ma(1+\qab)\Tat}\left(
\Tat-\Tar-\frac{\ma\qa\da\db\bm{\Omega}_i\cdot\bm{\Omega}_j}{4\dr}\right)}$\\
\vvss
$\Xi_{ij}^{(3)}$&&$\displaystyle{\frac{\nu_{ij}}{\dt}\frac{4\mab^2\qab^2(1+\eet)}{(1+\qab)^2\Tar}\left[\frac{\Tar-\Tbr}{\ma\mb\qa\qb}+
\frac{\Tat-\Tbt}{\ma\mb\qa}
+\frac{\Tar-\Tat}{\ma\mab\qa}
+\left(\frac{\ma\qa-\mb\qb}{\ma\mb\qa\qb}+\frac{1}{\mab}\right)\frac{\da\db\bm{\Omega}_i\cdot\bm{\Omega}_j}{4\dr}
\right]}$\\
\hline
\end{tabular}
\end{table}

Substitution of the expression for $\llangle \g\rrangle$ into Eq.\ \eqref{coll_freq} provides the approximate expression
\beq
\label{nuij}
\nu_{ij}=\frac{\sqrt{2}\pi^{(\dt-1)/2}}{\Gamma(\dt/2)}\chi_{ij}n_j\sigma_{ij}^{\dt-1}\left(\frac{\Tt_i}{m_i}+\frac{\Tt_j}{m_j}\right)^{1/2}.
\eeq
Analogously, substitution into Table \ref{table:1} of the expressions for other two-body averages shown in Table \ref{table:2} allows us to obtain the energy production rates $\zabt$ and $\zabr$ defined  by Eq.\ \eqref{prRates}. The results are given in Table \ref{table:3}, where the equality $\dr=\frac{1}{2}\dt(\dt-1)$ (valid only for disks and spheres) has been used in order to present the expressions in a compact form. In fact, we can observe that the number of degrees of freedom $\dt$ and $\dr$ intervene in $\zabt$ and $\zabr$ by following three simple rules: (i) $\zabt$ and $\zabr$ are divided by $\dt$ and $\dr$, respectively, as a consequence of their definitions in Eq.\ \eqref{prRates}; (ii) a factor $\dr/\dt$ is attached to $\ett$ and $\ett^2$; (iii) a factor $\dr^{-1}$ is attached to $\bm{\Omega}_\al \cdot \bm{\Omega}_\be$.

In the special case of frictionless, smooth particles ($\eet=-1\Rightarrow\ett=0$), one has $\zabr=0$ and
\beq
\label{zabt_smooth}
\zabt=\frac{2 \nu_{ij}\mab^2 }{\dt \ma \Tat }\left[{2\enn}\frac{\Tat}{\mab}-{\enn^2}\left(\frac{\Tat}{\ma}+\frac{\Tbt}{\mb}\right)\right].
\eeq
This coincides with previous results for an arbitrary number of dimensions $d=\dt$ \cite{G02,SA05,GM07,VGS07}. On the other hand, in the case of rough disks ($\dt=2$, $\dr=1$) or spheres ($\dt=3$, $\dr=3$), the expressions for $\zabt$ and $\zabr$ in Table \ref{table:3} reduce to results derived in Refs.\ \cite{S18} and \cite{SKG10}, respectively.

The global cooling rate $\zeta$ defined in Eq.\ \eqref{prRatesTotal} is also shown in Table \ref{table:3}. Now, a factor $\dr/\dt$ is attached to $(1-\eet^2)$. While $\zeta$ is a positive definite quantity, the partial production rates $\zabt$ and $\zabr$ can in general be positive or negative since the energy dissipation and equipartition effects are mixed together. To disentangle them, it is convenient to carry out the decompositions \cite{S11b,S18}
\beq
\label{B4}
\zabt=\frac{\dr\qa\Tar}{\dt\Tat}\zabr+\zeta_{\al\be}^\tr+\Xi_{ij}^{(1)}
+\Xi_{ij}^{(2)},\quad
\zabr=\zeta_{\al\be}^\rot+\Xi_{ij}^{(3)},
\eeq
where $\zeta_{\al\be}^\tr$ and $\zeta_{\al\be}^\rot$ are true cooling rates (positive definite), whereas $\Xi_{ij}^{(\text{1--3})}$ represent \emph{equipartition} rates and do not have a definite sign. The expressions for $\zeta_{\al\be}^\tr$, $\zeta_{\al\be}^\rot$, and $\Xi_{ij}^{(1\text{--}3)}$ are also included in Table \ref{table:3}.
It can be easily checked that ${\na}\Tat\Xi_{\al\be}^{(1)}+{\nb}\Tbt\Xi_{\be\al}^{(1)}=0$ and $\na\left[\dt\Tat\Xi_{\al\be}^{(2)}+\dr(1+\qa)\Tar\Xi_{\al\be}^{(3)}\right]+\nb\left[\dt\Tbt\Xi_{\be\al}^{(2)}+\dr(1+\qb)\Tbr\Xi_{\be\al}^{(3)}\right]=0$.
Therefore, as expected on physical grounds, the equipartition rates $\Xi_{ij}^{(1\text{--}3)}$ do not contribute to the net cooling rate $\zeta$, so that
\beq
\label{B8}
\zeta=\frac{1}{2(\dt+\dr)nT}\sum_{i,j}\left\{\na\left[\dt\Tat\zeta_{ij}^\tr+\dr({1+\qa})\Tar\zeta_{ij}^\rot\right]+\nb\left[\dt\Tbt\zeta_{ji}^\tr+\dr({1+\qb})\Tbr\zeta_{ji}^\rot\right]\right\}.
\eeq

Apart from the energy production rates shown in Table \ref{table:3}, one can introduce a spin production rate $\zeta_{ij}^\Omega$ by \cite{KSG14}
\beq
\frac{\sigma_i}{n_i}\mathcal{J}_{ij}[\bm{\omega}_i|f_i,f_j]=-\frac{\zeta_{ij}^\Omega}{2}\left(\sigma_i\bm{\Omega}_i+\sigma_j\bm{\Omega}_j\right),\quad \zeta_{ij}^{\Omega}=\frac{\nu_{ij}}{\dt}\frac{4\mab\ett}{\ma\qa},
\eeq
where in the last equality use has been made of Tables \ref{table:1} and \ref{table:2}.

\begin{table}[t]
\caption{Collisional energy production rates for monodisperse systems}
\label{table:4}
\tabcolsep7pt\begin{tabular}{rcl}
\hline
%Quantity && \hspace{5cm} Expression\\
%\hline
\vvss
$\xi^\tr$&&$\displaystyle{\frac{\nu}{\dt}\left\{1-\alpha^2+\frac{2\dr\kappa(1+\beta)}{\dt(1+\kappa)^2 T^\tr}\left[\frac{\kappa(1-\beta)}{2}\left(T^\tr+\frac{T^\rot}{\kappa}+\frac{m\sigma^2\Omega^2}{4\dr}\right)
+T^\tr-T^\rot-\frac{{\kappa}m\sigma^2\Omega^2}{4\dr}\right]\right\}}$\\
\vvss
$\xi^\rot$&&$\displaystyle{\frac{\nu}{\dt}\frac{2\kappa(1+\beta)}{(1+\kappa)^2T^\rot}\left[
\frac{1-\beta}{2}\left(T^\tr+\frac{T^\rot}{\kappa}+\frac{m\sigma^2\Omega^2}{4\dr}\right)
+T^\rot-T^\tr+\frac{{\kappa}m\sigma^2\Omega^2}{4\dr}\right]}$\\
\vvss
$\zeta$&&$\displaystyle{\frac{\nu}{(\dt+\dr)T}\left[\left(1-\alpha^2\right)T^\tr+ \frac{\dr\kappa(1-\beta^2)}{\dt(1+\kappa)}\left(T^\tr+\frac{T^\rot}{\kappa}+\frac{m\sigma^2\Omega^2}{4\dr}\right)\right]}$\\
\vvss
$\zeta^\tr$&&$\displaystyle{\frac{\nu}{\dt}\left(1-\alpha^2\right)}$\\
\vvss
$\zeta^\rot$&&$\displaystyle{\frac{\nu}{\dt}\frac{\kappa(1-\beta^2)}{(1+\kappa)^2T^\rot}\left(T^\tr+
\frac{T^\rot}{\kappa}+\frac{m\sigma^2\Omega^2}{4\dr}\right)}$\\
\vvss
$\Xi^{(1)}$&&$0$\\
\vvss
$\Xi^{(2)}$&&$\displaystyle{\frac{\nu}{\dt^2}\frac{2\dr\kappa(1+\beta)}{(1+\kappa)T^\tr}\left(T^\tr-T^\rot-\frac{{\kappa}m\sigma^2\Omega^2}{4\dr}\right)}$\\
\vvss
$\Xi^{(3)}$&&$\displaystyle{-\frac{\dt}{\dr(1+\kappa)}\frac{T^\tr}{T^\rot}\Xi^{(2)}}$\\
\hline
\end{tabular}
\end{table}

The expressions of Table \ref{table:3} simplify considerably in the case of monodisperse systems, i.e., if $\ma= 2\mab=m$, $\qa=\qab= \kappa$, $\da= \dab=\sigma$, $\een=\alpha$, $\eet=\beta$, $\chi_{ij}=\chi$, $\Tat=\Tt$, $\Tar=\Tr$,  $\bm{\Omega}_{\al}= \bm{\Omega}$, and $\nu_{ij}=\nu$, with
\beq
\nu=\frac{2\pi^{(\dt-1)/2}}{\Gamma(\dt/2)}\chi n\sigma^{\dt-1}\sqrt{\frac{\Tt}{m}},
\eeq
for all $i$ and $j$. From those conditions, Table \ref{table:3} reduces to Table \ref{table:4}. Moreover, the spin production rate becomes $\zeta_{ij}^\Omega=\zeta^\Omega=2\nu(1+\beta)/\dt(1+\kappa)$.

\section{CONCLUSION}

Arguably, the most distinctive feature of granular gases is collisional energy dissipation due to inelasticity and surface roughness of the particles. Moreover, there are in general two classes of contributions to the kinetic energy, one associated with $\dt$ translational degrees of freedom and the other one associated with $\dr$ rotational degrees of freedom. In a multicomponent gas, additionally, the (translational or rotational) kinetic energy is split into different components. To characterize all these separate contributions, the (partial) granular temperatures $\Tat$ and $\Tar$ are defined as twice the mean (translational or rotational) kinetic energy per particle and per degree of freedom associated with component $i$. Collisions of particles of component $i$ with those of component $j$ produce two main competing effects: on the one hand, $\Tat$, $\Tbt$, $\Tar$, and $\Tbr$ tend to decay due to a dissipative cooling effect but, on the other hand,  those partial temperatures also tend to equal each other due to an equipartition effect. These basic effects are entangled in the energy production rates $\zabt$ and $\zabr$ defined by the rate equations $\left.\partial_t\Tat\right|_{\text{coll},j}=-\zabt\Tat$ and $\left.\partial_t\Tar\right|_{\text{coll},j}=-\zabr\Tar$.

The aim of this work has been the unified  derivation of $\zabt$ and $\zabr$ for disks ($\dt=2$, $\dr=1$) and spheres ($\dt=\dr=3$) from the Boltzmann equation (i.e., under the molecular chaos ansatz). In order to obtain analytic results, statistical independence of the distributions of translational and angular velocities and a Maxwellian form for the translational distribution have been assumed. The expressions for $\zabt$ and $\zabr$, together with those for the global cooling rate $\zeta$, the partial cooling rates $\zeta_{ij}^\tr$ and $\zeta_{ij}^\rot$, and the equipartition rates $\Xi_{ij}^{(1)\text{--}(3)}$, are presented in Table \ref{table:3}. They encapsulate isolated previous results for smooth $d$-dimensional spheres \cite{G02,SA05,GM07,VGS07}, rough disks \cite{S18}, and rough spheres \cite{SKG10} in a common framework. The parametric dependence of $\zabt$ and $\zabr$ on the numbers of degrees of freedom $\dt$ and $\dr$ turns out to be quite simple: $\dt\zabt/\nu_{ij}$ and $\dr\zabr/\nu_{ij}$, where the collision frequency $\nu_{ij}$ depends on $\dt$ [see Eq.\ \eqref{nuij}], have a factor $\dr/\dt$ attached to $\ett$ and $\ett^2$, and a factor $\dr^{-1}$ attached to $\bm{\Omega}_i\cdot\bm{\Omega}_j$.

%As an immediate application of the results reported in this work, we plan to carry out a unified study (i.e., in terms of $\dt$ and $\dr$) of homogeneous states, both undriven (homogeneous cooling state) and driven (stochastic thermostat), together with a critical comparison between the behaviors of rough disks and rough spheres. As a next logical step, the Navier--Stokes hydrodynamic equations, already known for rough spheres \cite{KSG14}, will be derived in terms of $\dt$ and $\dr$.

% Sections that will go in second font

% Acknowledgement
\section{ACKNOWLEDGMENTS}
A.M. is grateful to the Ministerio de Educaci\'on, Cultura y Deporte (Spain) for a Beca-Colaboraci\'on during the academic year 2017--2018. The research of A.S. has been supported by the Spanish Agencia Estatal de Investigaci\'on through Grant No.\ FIS2016-76359-P
and the Junta de Extremadura (Spain) through Grant No.\ GR18079, both partially financed by Fondo Europeo de
Desarrollo Regional funds.

% References

\nocite{*}
%\bibliographystyle{aipnum-cp}%
%\bibliography{D:/Dropbox/Mis_Dropcumentos/bib_files/Granular}%

\begin{thebibliography}{410}%
\makeatletter
\providecommand \@ifxundefined [1]{%
 \@ifx{#1\undefined}
}%
\providecommand \@ifnum [1]{%
 \ifnum #1\expandafter \@firstoftwo
 \else \expandafter \@secondoftwo
 \fi
}%
\providecommand \@ifx [1]{%
 \ifx #1\expandafter \@firstoftwo
 \else \expandafter \@secondoftwo
 \fi
}%
\providecommand \natexlab [1]{#1}%
\providecommand \enquote  [1]{``#1''}%
\providecommand \bibnamefont  [1]{#1}%
\providecommand \bibfnamefont [1]{#1}%
\providecommand \citenamefont [1]{#1}%
\providecommand \href@noop [0]{\@secondoftwo}%
\providecommand \href [0]{\begingroup \@sanitize@url \@href}%
\providecommand \@href[1]{\@@startlink{#1}\@@href}%
\providecommand \@@href[1]{\endgroup#1\@@endlink}%
\providecommand \@sanitize@url [0]{\catcode `\$12\catcode `\&12\catcode
  `\#12\catcode `\^12\catcode `\_12\catcode `\%12\relax}%
\providecommand \@@startlink[1]{}%
\providecommand \@@endlink[0]{}%
\providecommand \url  [0]{\begingroup\@sanitize@url \@url }%
\providecommand \@url [1]{\endgroup\@href {#1}{\urlprefix }}%
\providecommand \urlprefix  [0]{URL }%
\providecommand \Eprint [0]{\href }%
\providecommand \doibase [0]{http://dx.doi.org/}%
\providecommand \selectlanguage [0]{\@gobble}%
\providecommand \bibinfo  [0]{\@secondoftwo}%
\providecommand \bibfield  [0]{\@secondoftwo}%
\providecommand \translation [1]{[#1]}%
\providecommand \BibitemOpen [0]{}%
\providecommand \bibitemStop [0]{}%
\providecommand \bibitemNoStop [0]{.\EOS\space}%
\providecommand \EOS [0]{\spacefactor3000\relax}%
\providecommand \BibitemShut  [1]{\csname bibitem#1\endcsname}%
\let\auto@bib@innerbib\@empty
%</preamble>
\bibitem [{\citenamefont {Dufty}(2000)}]{D00}%
  \BibitemOpen
  \bibfield  {author} {\bibinfo {author} {\bibfnamefont {J.~W.}\ \bibnamefont
  {Dufty}},\ }\href {\doibase 10.1088/0953-8984/12/8A/306} {\bibfield
  {journal} {\bibinfo  {journal} {J. Phys.: Condens. Matter}\ }\textbf
  {\bibinfo {volume} {12}},\ \unskip\ \bibinfo {pages} {A47--A56} (\bibinfo
  {year} {2000})}\BibitemShut {NoStop}%
\bibitem [{\citenamefont {Ottino}\ and\ \citenamefont {Khakhar}(2000)}]{OK00}%
  \BibitemOpen
  \bibfield  {author} {\bibinfo {author} {\bibfnamefont {J.~M.}\ \bibnamefont
  {Ottino}}\ and\ \bibinfo {author} {\bibfnamefont {D.~V.}\ \bibnamefont
  {Khakhar}},\ }\href {\doibase 10.1146/annurev.fluid.32.1.55} {\bibfield
  {journal} {\bibinfo  {journal} {Annu. Rev. Fluid Mech.}\ }\textbf {\bibinfo
  {volume} {32}},\ \unskip\ \bibinfo {pages} {55--91} (\bibinfo {year}
  {2000})}\BibitemShut {NoStop}%
\bibitem [{\citenamefont {P\"oschel}\ and\ \citenamefont
  {Luding}(2001)}]{PL01}%
  \BibitemOpen
  \bibinfo {editor} {\bibfnamefont {T.}~\bibnamefont {P\"oschel}}\ and\
  \bibinfo {editor} {\bibfnamefont {S.}~\bibnamefont {Luding}},\ eds.,\
  \href@noop {} {\emph {\bibinfo {title} {Granular Gases}}},\ \bibinfo {series}
  {Lecture Notes in Physics}, Vol.\ \bibinfo {volume} {564}\ (\bibinfo
  {publisher} {Springer},\ \bibinfo {address} {Berlin},\ \bibinfo {year}
  {2001})\BibitemShut {NoStop}%
\bibitem [{\citenamefont {Goldhirsch}(2003)}]{G03}%
  \BibitemOpen
  \bibfield  {author} {\bibinfo {author} {\bibfnamefont {I.}~\bibnamefont
  {Goldhirsch}},\ }\href {\doibase 10.1146/annurev.fluid.35.101101.161114}
  {\bibfield  {journal} {\bibinfo  {journal} {Annu. Rev. Fluid Mech.}\ }\textbf
  {\bibinfo {volume} {35}},\ \unskip\ \bibinfo {pages} {267--293} (\bibinfo
  {year} {2003})}\BibitemShut {NoStop}%
\bibitem [{\citenamefont {Kudrolli}(2004)}]{K04}%
  \BibitemOpen
  \bibfield  {author} {\bibinfo {author} {\bibfnamefont {A.}~\bibnamefont
  {Kudrolli}},\ }\href {\doibase 10.1088/0034-4885/67/3/R01} {\bibfield
  {journal} {\bibinfo  {journal} {Rep. Prog. Phys.}\ }\textbf {\bibinfo
  {volume} {67}},\ \unskip\ \bibinfo {pages} {209--247} (\bibinfo {year}
  {2004})}\BibitemShut {NoStop}%
\bibitem [{\citenamefont {Brilliantov}\ and\ \citenamefont
  {P\"oschel}(2004)}]{BP04}%
  \BibitemOpen
  \bibfield  {author} {\bibinfo {author} {\bibfnamefont {N.~V.}\ \bibnamefont
  {Brilliantov}}\ and\ \bibinfo {author} {\bibfnamefont {T.}~\bibnamefont
  {P\"oschel}},\ }\href@noop {} {\emph {\bibinfo {title} {Kinetic Theory of
  Granular Gases}}}\ (\bibinfo  {publisher} {Oxford University Press},\
  \bibinfo {address} {Oxford},\ \bibinfo {year} {2004})\BibitemShut {NoStop}%
\bibitem [{\citenamefont {Aranson}\ and\ \citenamefont
  {Tsimring}(2006)}]{AT06}%
  \BibitemOpen
  \bibfield  {author} {\bibinfo {author} {\bibfnamefont {I.~S.}\ \bibnamefont
  {Aranson}}\ and\ \bibinfo {author} {\bibfnamefont {L.~S.}\ \bibnamefont
  {Tsimring}},\ }\href {\doibase 10.1103/RevModPhys.78.641} {\bibfield
  {journal} {\bibinfo  {journal} {Rev. Mod. Phys.}\ }\textbf {\bibinfo {volume}
  {78}},\ \unskip\ \bibinfo {pages} {641--692} (\bibinfo {year}
  {2006})}\BibitemShut {NoStop}%
\bibitem [{\citenamefont {Rao}\ and\ \citenamefont {Nott}(2008)}]{RN08}%
  \BibitemOpen
  \bibfield  {author} {\bibinfo {author} {\bibfnamefont {K.~K.}\ \bibnamefont
  {Rao}}\ and\ \bibinfo {author} {\bibfnamefont {P.~R.}\ \bibnamefont {Nott}},\
  }\href@noop {} {\emph {\bibinfo {title} {An Introduction to Granular Flow}}}\
  (\bibinfo  {publisher} {Cambridge University Press},\ \bibinfo {address}
  {Cambridge, England},\ \bibinfo {year} {2008})\BibitemShut {NoStop}%
\bibitem [{\citenamefont {Jenkins}\ and\ \citenamefont
  {Richman}(1985{\natexlab{a}})}]{JR85a}%
  \BibitemOpen
  \bibfield  {author} {\bibinfo {author} {\bibfnamefont {J.~T.}\ \bibnamefont
  {Jenkins}}\ and\ \bibinfo {author} {\bibfnamefont {M.~W.}\ \bibnamefont
  {Richman}},\ }\href {\doibase 10.1063/1.865302} {\bibfield  {journal}
  {\bibinfo  {journal} {Phys. Fluids}\ }\textbf {\bibinfo {volume} {28}},\
  \unskip\ \bibinfo {pages} {3485--3494} (\bibinfo {year}
  {1985}{\natexlab{a}})}\BibitemShut {NoStop}%
\bibitem [{\citenamefont {Lun}\ and\ \citenamefont {Savage}(1987)}]{LS87}%
  \BibitemOpen
  \bibfield  {author} {\bibinfo {author} {\bibfnamefont {C.~K.~K.}\
  \bibnamefont {Lun}}\ and\ \bibinfo {author} {\bibfnamefont {S.~B.}\
  \bibnamefont {Savage}},\ }\href {\doibase 10.1115/1.3172993} {\bibfield
  {journal} {\bibinfo  {journal} {J. Appl. Mech.}\ }\textbf {\bibinfo {volume}
  {54}},\ \unskip\ \bibinfo {pages} {47--53} (\bibinfo {year}
  {1987})}\BibitemShut {NoStop}%
\bibitem [{\citenamefont {Campbell}(1989)}]{C89}%
  \BibitemOpen
  \bibfield  {author} {\bibinfo {author} {\bibfnamefont {C.~S.}\ \bibnamefont
  {Campbell}},\ }\href {\doibase 10.1017/S0022112089001540} {\bibfield
  {journal} {\bibinfo  {journal} {J. Fluid Mech.}\ }\textbf {\bibinfo {volume}
  {203}},\ \unskip\ \bibinfo {pages} {449--473} (\bibinfo {year}
  {1989})}\BibitemShut {NoStop}%
\bibitem [{\citenamefont {Lun}(1991)}]{L91}%
  \BibitemOpen
  \bibfield  {author} {\bibinfo {author} {\bibfnamefont {C.~K.~K.}\
  \bibnamefont {Lun}},\ }\href {\doibase 10.1017/S0022112091000599} {\bibfield
  {journal} {\bibinfo  {journal} {J. Fluid Mech.}\ }\textbf {\bibinfo {volume}
  {233}},\ \unskip\ \bibinfo {pages} {539--559} (\bibinfo {year}
  {1991})}\BibitemShut {NoStop}%
\bibitem [{\citenamefont {Lun}\ and\ \citenamefont {Bent}(1994)}]{LB94}%
  \BibitemOpen
  \bibfield  {author} {\bibinfo {author} {\bibfnamefont {C.~K.~K.}\
  \bibnamefont {Lun}}\ and\ \bibinfo {author} {\bibfnamefont {A.~A.}\
  \bibnamefont {Bent}},\ }\href {\doibase 10.1017/S0022112094003356} {\bibfield
   {journal} {\bibinfo  {journal} {J. Fluid Mech.}\ }\textbf {\bibinfo {volume}
  {258}},\ \unskip\ \bibinfo {pages} {335--353} (\bibinfo {year}
  {1994})}\BibitemShut {NoStop}%
\bibitem [{\citenamefont {Goldshtein}\ and\ \citenamefont
  {Shapiro}(1995)}]{GS95}%
  \BibitemOpen
  \bibfield  {author} {\bibinfo {author} {\bibfnamefont {A.}~\bibnamefont
  {Goldshtein}}\ and\ \bibinfo {author} {\bibfnamefont {M.}~\bibnamefont
  {Shapiro}},\ }\href {\doibase 10.1017/S0022112095000048} {\bibfield
  {journal} {\bibinfo  {journal} {J. Fluid Mech.}\ }\textbf {\bibinfo {volume}
  {282}},\ \unskip\ \bibinfo {pages} {75--114} (\bibinfo {year}
  {1995})}\BibitemShut {NoStop}%
\bibitem [{\citenamefont {Luding}(1995)}]{L95}%
  \BibitemOpen
  \bibfield  {author} {\bibinfo {author} {\bibfnamefont {S.}~\bibnamefont
  {Luding}},\ }\href {\doibase 10.1103/PhysRevE.52.4442} {\bibfield  {journal}
  {\bibinfo  {journal} {Phys. Rev. E}\ }\textbf {\bibinfo {volume} {52}},\
  \unskip\ \bibinfo {pages} {4442--4457} (\bibinfo {year} {1995})}\BibitemShut
  {NoStop}%
\bibitem [{\citenamefont {Lun}(1996)}]{L96}%
  \BibitemOpen
  \bibfield  {author} {\bibinfo {author} {\bibfnamefont {C.~K.~K.}\
  \bibnamefont {Lun}},\ }\href {\doibase 10.1063/1.869068} {\bibfield
  {journal} {\bibinfo  {journal} {Phys. Fluids}\ }\textbf {\bibinfo {volume}
  {8}},\ \unskip\ \bibinfo {pages} {2868--2883} (\bibinfo {year}
  {1996})}\BibitemShut {NoStop}%
\bibitem [{\citenamefont {Zamankhan}\ \emph {et~al.}(1998)\citenamefont
  {Zamankhan}, \citenamefont {Tafreshi}, \citenamefont {Polashenski},
  \citenamefont {Sarkomaa},\ and\ \citenamefont {Hyndman}}]{ZTPSH98}%
  \BibitemOpen
  \bibfield  {author} {\bibinfo {author} {\bibfnamefont {P.}~\bibnamefont
  {Zamankhan}}, \bibinfo {author} {\bibfnamefont {H.~V.}\ \bibnamefont
  {Tafreshi}}, \bibinfo {author} {\bibfnamefont {W.}~\bibnamefont
  {Polashenski}}, \bibinfo {author} {\bibfnamefont {P.}~\bibnamefont
  {Sarkomaa}}, \ and\ \bibinfo {author} {\bibfnamefont {C.~L.}\ \bibnamefont
  {Hyndman}},\ }\href {\doibase 10.1063/1.477076} {\bibfield  {journal}
  {\bibinfo  {journal} {J. Chem. Phys.}\ }\textbf {\bibinfo {volume} {109}},\
  \unskip\ \bibinfo {pages} {4487--4491} (\bibinfo {year} {1998})}\BibitemShut
  {NoStop}%
\bibitem [{\citenamefont {Huthmann}\ and\ \citenamefont
  {Zippelius}(1997)}]{HZ97}%
  \BibitemOpen
  \bibfield  {author} {\bibinfo {author} {\bibfnamefont {M.}~\bibnamefont
  {Huthmann}}\ and\ \bibinfo {author} {\bibfnamefont {A.}~\bibnamefont
  {Zippelius}},\ }\href {\doibase 10.1103/PhysRevE.56.R6275} {\bibfield
  {journal} {\bibinfo  {journal} {Phys. Rev. E}\ }\textbf {\bibinfo {volume}
  {56}},\ \unskip\ \bibinfo {pages} {R6275--R6278} (\bibinfo {year}
  {1997})}\BibitemShut {NoStop}%
\bibitem [{\citenamefont {McNamara}\ and\ \citenamefont {Luding}(1998)}]{ML98}%
  \BibitemOpen
  \bibfield  {author} {\bibinfo {author} {\bibfnamefont {S.}~\bibnamefont
  {McNamara}}\ and\ \bibinfo {author} {\bibfnamefont {S.}~\bibnamefont
  {Luding}},\ }\href {\doibase 10.1103/PhysRevE.58.2247} {\bibfield  {journal}
  {\bibinfo  {journal} {Phys. Rev. E}\ }\textbf {\bibinfo {volume} {58}},\
  \unskip\ \bibinfo {pages} {2247--2250} (\bibinfo {year} {1998})}\BibitemShut
  {NoStop}%
\bibitem [{\citenamefont {Luding}\ \emph {et~al.}(1998)\citenamefont {Luding},
  \citenamefont {Huthmann}, \citenamefont {McNamara},\ and\ \citenamefont
  {Zippelius}}]{LHMZ98}%
  \BibitemOpen
  \bibfield  {author} {\bibinfo {author} {\bibfnamefont {S.}~\bibnamefont
  {Luding}}, \bibinfo {author} {\bibfnamefont {M.}~\bibnamefont {Huthmann}},
  \bibinfo {author} {\bibfnamefont {S.}~\bibnamefont {McNamara}}, \ and\
  \bibinfo {author} {\bibfnamefont {A.}~\bibnamefont {Zippelius}},\ }\href
  {\doibase 10.1103/PhysRevE.58.3416} {\bibfield  {journal} {\bibinfo
  {journal} {Phys. Rev. E}\ }\textbf {\bibinfo {volume} {58}},\ \unskip\
  \bibinfo {pages} {3416--3425} (\bibinfo {year} {1998})}\BibitemShut {NoStop}%
\bibitem [{\citenamefont {Herbst}, \citenamefont {Huthmann},\ and\
  \citenamefont {Zippelius}(2000)}]{HHZ00}%
  \BibitemOpen
  \bibfield  {author} {\bibinfo {author} {\bibfnamefont {O.}~\bibnamefont
  {Herbst}}, \bibinfo {author} {\bibfnamefont {M.}~\bibnamefont {Huthmann}}, \
  and\ \bibinfo {author} {\bibfnamefont {A.}~\bibnamefont {Zippelius}},\ }\href
  {\doibase 10.1007/PL00010915} {\bibfield  {journal} {\bibinfo  {journal}
  {Granul. Matter}\ }\textbf {\bibinfo {volume} {2}},\ \unskip\ \bibinfo
  {pages} {211--219} (\bibinfo {year} {2000})}\BibitemShut {NoStop}%
\bibitem [{\citenamefont {Aspelmeier}, \citenamefont {Huthmann},\ and\
  \citenamefont {Zippelius}(2001)}]{AHZ01}%
  \BibitemOpen
  \bibfield  {author} {\bibinfo {author} {\bibfnamefont {T.}~\bibnamefont
  {Aspelmeier}}, \bibinfo {author} {\bibfnamefont {M.}~\bibnamefont
  {Huthmann}}, \ and\ \bibinfo {author} {\bibfnamefont {A.}~\bibnamefont
  {Zippelius}},\ }\enquote {\bibinfo {title} {Free cooling of particles with
  rotational degrees of freedom},}\ in\ \href@noop {} {\emph {\bibinfo
  {booktitle} {Granular Gases}}},\ \bibinfo {series} {Lectures Notes in
  Physics}, Vol.\ \bibinfo {volume} {564},\ \bibinfo {editor} {edited by\
  \bibinfo {editor} {\bibfnamefont {T.}~\bibnamefont {P\"oschel}}\ and\
  \bibinfo {editor} {\bibfnamefont {S.}~\bibnamefont {Luding}}}\ (\bibinfo
  {publisher} {Springer},\ \bibinfo {address} {Berlin},\ \bibinfo {year}
  {2001})\ \unskip, pp.\ \bibinfo {pages} {31--58}\BibitemShut {NoStop}%
\bibitem [{\citenamefont {Mitarai}, \citenamefont {Hayakawa},\ and\
  \citenamefont {Nakanishi}(2002)}]{MHN02}%
  \BibitemOpen
  \bibfield  {author} {\bibinfo {author} {\bibfnamefont {N.}~\bibnamefont
  {Mitarai}}, \bibinfo {author} {\bibfnamefont {H.}~\bibnamefont {Hayakawa}}, \
  and\ \bibinfo {author} {\bibfnamefont {H.}~\bibnamefont {Nakanishi}},\ }\href
  {\doibase 10.1103/PhysRevLett.88.174301} {\bibfield  {journal} {\bibinfo
  {journal} {Phys. Rev. Lett.}\ }\textbf {\bibinfo {volume} {88}},\ p.\
  \bibinfo {pages} {174301} (\bibinfo {year} {2002})}\BibitemShut {NoStop}%
\bibitem [{\citenamefont {Cafiero}, \citenamefont {Luding},\ and\ \citenamefont
  {Herrmann}(2002)}]{CLH02}%
  \BibitemOpen
  \bibfield  {author} {\bibinfo {author} {\bibfnamefont {R.}~\bibnamefont
  {Cafiero}}, \bibinfo {author} {\bibfnamefont {S.}~\bibnamefont {Luding}}, \
  and\ \bibinfo {author} {\bibfnamefont {H.~J.}\ \bibnamefont {Herrmann}},\
  }\href {\doibase 10.1209/epl/i2002-00295-7} {\bibfield  {journal} {\bibinfo
  {journal} {Europhys. Lett.}\ }\textbf {\bibinfo {volume} {60}},\ \unskip\
  \bibinfo {pages} {854--860} (\bibinfo {year} {2002})}\BibitemShut {NoStop}%
\bibitem [{\citenamefont {Jenkins}\ and\ \citenamefont {Zhang}(2002)}]{JZ02}%
  \BibitemOpen
  \bibfield  {author} {\bibinfo {author} {\bibfnamefont {J.~T.}\ \bibnamefont
  {Jenkins}}\ and\ \bibinfo {author} {\bibfnamefont {C.}~\bibnamefont
  {Zhang}},\ }\href {\doibase 10.1063/1.1449466} {\bibfield  {journal}
  {\bibinfo  {journal} {Phys. Fluids}\ }\textbf {\bibinfo {volume} {14}},\
  \unskip\ \bibinfo {pages} {1228--1235} (\bibinfo {year} {2002})}\BibitemShut
  {NoStop}%
\bibitem [{\citenamefont {Polashenski}\ \emph {et~al.}(2002)\citenamefont
  {Polashenski}, \citenamefont {Zamankhan}, \citenamefont {M\"akiharju},\ and\
  \citenamefont {Zamankhan}}]{PZMZ02}%
  \BibitemOpen
  \bibfield  {author} {\bibinfo {author} {\bibfnamefont {W.}~\bibnamefont
  {Polashenski}}, \bibinfo {author} {\bibfnamefont {P.}~\bibnamefont
  {Zamankhan}}, \bibinfo {author} {\bibfnamefont {S.}~\bibnamefont
  {M\"akiharju}}, \ and\ \bibinfo {author} {\bibfnamefont {P.}~\bibnamefont
  {Zamankhan}},\ }\href {\doibase 10.1103/PhysRevE.66.021303} {\bibfield
  {journal} {\bibinfo  {journal} {Phys. Rev. E}\ }\textbf {\bibinfo {volume}
  {66}},\ p.\ \bibinfo {pages} {021303} (\bibinfo {year} {2002})}\BibitemShut
  {NoStop}%
\bibitem [{\citenamefont {Moon}, \citenamefont {Swift},\ and\ \citenamefont
  {Swinney}(2004)}]{MSS04}%
  \BibitemOpen
  \bibfield  {author} {\bibinfo {author} {\bibfnamefont {S.~J.}\ \bibnamefont
  {Moon}}, \bibinfo {author} {\bibfnamefont {J.~B.}\ \bibnamefont {Swift}}, \
  and\ \bibinfo {author} {\bibfnamefont {H.~L.}\ \bibnamefont {Swinney}},\
  }\href {\doibase 10.1103/PhysRevE.69.031301} {\bibfield  {journal} {\bibinfo
  {journal} {Phys. Rev. E}\ }\textbf {\bibinfo {volume} {69}},\ p.\ \bibinfo
  {pages} {031301} (\bibinfo {year} {2004})}\BibitemShut {NoStop}%
\bibitem [{\citenamefont {Herbst}\ \emph {et~al.}(2005)\citenamefont {Herbst},
  \citenamefont {Cafiero}, \citenamefont {Zippelius}, \citenamefont
  {Herrmann},\ and\ \citenamefont {Luding}}]{HCZHL05}%
  \BibitemOpen
  \bibfield  {author} {\bibinfo {author} {\bibfnamefont {O.}~\bibnamefont
  {Herbst}}, \bibinfo {author} {\bibfnamefont {R.}~\bibnamefont {Cafiero}},
  \bibinfo {author} {\bibfnamefont {A.}~\bibnamefont {Zippelius}}, \bibinfo
  {author} {\bibfnamefont {H.~J.}\ \bibnamefont {Herrmann}}, \ and\ \bibinfo
  {author} {\bibfnamefont {S.}~\bibnamefont {Luding}},\ }\href {\doibase
  10.1063/1.2049277} {\bibfield  {journal} {\bibinfo  {journal} {Phys. Fluids}\
  }\textbf {\bibinfo {volume} {17}},\ p.\ \bibinfo {pages} {107102} (\bibinfo
  {year} {2005})}\BibitemShut {NoStop}%
\bibitem [{\citenamefont {Goldhirsch}, \citenamefont {Noskowicz},\ and\
  \citenamefont {Bar-Lev}(2005{\natexlab{a}})}]{GNB05}%
  \BibitemOpen
  \bibfield  {author} {\bibinfo {author} {\bibfnamefont {I.}~\bibnamefont
  {Goldhirsch}}, \bibinfo {author} {\bibfnamefont {S.~H.}\ \bibnamefont
  {Noskowicz}}, \ and\ \bibinfo {author} {\bibfnamefont {O.}~\bibnamefont
  {Bar-Lev}},\ }\href {\doibase 10.1103/PhysRevLett.95.068002} {\bibfield
  {journal} {\bibinfo  {journal} {Phys. Rev. Lett.}\ }\textbf {\bibinfo
  {volume} {95}},\ p.\ \bibinfo {pages} {068002} (\bibinfo {year}
  {2005}{\natexlab{a}})}\BibitemShut {NoStop}%
\bibitem [{\citenamefont {Zippelius}(2006)}]{Z06}%
  \BibitemOpen
  \bibfield  {author} {\bibinfo {author} {\bibfnamefont {A.}~\bibnamefont
  {Zippelius}},\ }\href {\doibase 10.1016/j.physa.2006.04.012} {\bibfield
  {journal} {\bibinfo  {journal} {Physica A}\ }\textbf {\bibinfo {volume}
  {369}},\ \unskip\ \bibinfo {pages} {143--158} (\bibinfo {year}
  {2006})}\BibitemShut {NoStop}%
\bibitem [{\citenamefont {Brilliantov}\ \emph {et~al.}(2007)\citenamefont
  {Brilliantov}, \citenamefont {P\"oschel}, \citenamefont {Kranz},\ and\
  \citenamefont {Zippelius}}]{BPKZ07}%
  \BibitemOpen
  \bibfield  {author} {\bibinfo {author} {\bibfnamefont {N.~V.}\ \bibnamefont
  {Brilliantov}}, \bibinfo {author} {\bibfnamefont {T.}~\bibnamefont
  {P\"oschel}}, \bibinfo {author} {\bibfnamefont {W.~T.}\ \bibnamefont
  {Kranz}}, \ and\ \bibinfo {author} {\bibfnamefont {A.}~\bibnamefont
  {Zippelius}},\ }\href {\doibase 10.1103/PhysRevLett.98.128001} {\bibfield
  {journal} {\bibinfo  {journal} {Phys. Rev. Lett.}\ }\textbf {\bibinfo
  {volume} {98}},\ p.\ \bibinfo {pages} {128001} (\bibinfo {year}
  {2007})}\BibitemShut {NoStop}%
\bibitem [{\citenamefont {Gayen}\ and\ \citenamefont {Alam}(2008)}]{GA08}%
  \BibitemOpen
  \bibfield  {author} {\bibinfo {author} {\bibfnamefont {B.}~\bibnamefont
  {Gayen}}\ and\ \bibinfo {author} {\bibfnamefont {M.}~\bibnamefont {Alam}},\
  }\href {\doibase 10.1103/PhysRevLett.100.068002} {\bibfield  {journal}
  {\bibinfo  {journal} {Phys. Rev. Lett.}\ }\textbf {\bibinfo {volume} {100}},\
  p.\ \bibinfo {pages} {068002} (\bibinfo {year} {2008})}\BibitemShut {NoStop}%
\bibitem [{\citenamefont {Kranz}\ \emph {et~al.}(2009)\citenamefont {Kranz},
  \citenamefont {Brilliantov}, \citenamefont {P\"oschel},\ and\ \citenamefont
  {Zippelius}}]{KBPZ09}%
  \BibitemOpen
  \bibfield  {author} {\bibinfo {author} {\bibfnamefont {W.~T.}\ \bibnamefont
  {Kranz}}, \bibinfo {author} {\bibfnamefont {N.~V.}\ \bibnamefont
  {Brilliantov}}, \bibinfo {author} {\bibfnamefont {T.}~\bibnamefont
  {P\"oschel}}, \ and\ \bibinfo {author} {\bibfnamefont {A.}~\bibnamefont
  {Zippelius}},\ }\href {\doibase 10.1140/epjst/e2010-01196-0} {\bibfield
  {journal} {\bibinfo  {journal} {Eur. Phys. J. Spec. Top.}\ }\textbf {\bibinfo
  {volume} {179}},\ \unskip\ \bibinfo {pages} {91--111} (\bibinfo {year}
  {2009})}\BibitemShut {NoStop}%
\bibitem [{\citenamefont {Kremer}(2010{\natexlab{a}})}]{K10a}%
  \BibitemOpen
  \bibfield  {author} {\bibinfo {author} {\bibfnamefont {G.~M.}\ \bibnamefont
  {Kremer}},\ }\href@noop {} {\emph {\bibinfo {title} {An Introduction to the
  Boltzmann Equation and Transport Processes in Gases}}}\ (\bibinfo
  {publisher} {Springer},\ \bibinfo {address} {Berlin},\ \bibinfo {year}
  {2010})\BibitemShut {NoStop}%
\bibitem [{\citenamefont {Santos}(2011{\natexlab{a}})}]{S11a}%
  \BibitemOpen
  \bibfield  {author} {\bibinfo {author} {\bibfnamefont {A.}~\bibnamefont
  {Santos}},\ }\href {\doibase 10.1063/1.3562623} {\bibfield  {journal}
  {\bibinfo  {journal} {AIP Conf. Proc.}\ }\textbf {\bibinfo {volume} {1333}},\
  \unskip\ \bibinfo {pages} {41--48} (\bibinfo {year}
  {2011}{\natexlab{a}})}\BibitemShut {NoStop}%
\bibitem [{\citenamefont {Santos}, \citenamefont {Kremer},\ and\ \citenamefont
  {dos Santos}(2011)}]{SKS11}%
  \BibitemOpen
  \bibfield  {author} {\bibinfo {author} {\bibfnamefont {A.}~\bibnamefont
  {Santos}}, \bibinfo {author} {\bibfnamefont {G.~M.}\ \bibnamefont {Kremer}},
  \ and\ \bibinfo {author} {\bibfnamefont {M.}~\bibnamefont {dos Santos}},\
  }\href {\doibase 10.1063/1.3558876} {\bibfield  {journal} {\bibinfo
  {journal} {Phys. Fluids}\ }\textbf {\bibinfo {volume} {23}},\ p.\ \bibinfo
  {pages} {030604} (\bibinfo {year} {2011})}\BibitemShut {NoStop}%
\bibitem [{\citenamefont {Santos}\ and\ \citenamefont {Kremer}(2012)}]{SK12}%
  \BibitemOpen
  \bibfield  {author} {\bibinfo {author} {\bibfnamefont {A.}~\bibnamefont
  {Santos}}\ and\ \bibinfo {author} {\bibfnamefont {G.~M.}\ \bibnamefont
  {Kremer}},\ }\href {\doibase 10.1063/1.4769657} {\bibfield  {journal}
  {\bibinfo  {journal} {AIP Conf. Proc.}\ }\textbf {\bibinfo {volume} {1501}},\
  \unskip\ \bibinfo {pages} {1044--1050} (\bibinfo {year} {2012})}\BibitemShut
  {NoStop}%
\bibitem [{\citenamefont {Mitrano}\ \emph {et~al.}(2013)\citenamefont
  {Mitrano}, \citenamefont {Dahl}, \citenamefont {Hilger}, \citenamefont
  {Ewasko},\ and\ \citenamefont {Hrenya}}]{MDHEH13}%
  \BibitemOpen
  \bibfield  {author} {\bibinfo {author} {\bibfnamefont {P.~P.}\ \bibnamefont
  {Mitrano}}, \bibinfo {author} {\bibfnamefont {S.~R.}\ \bibnamefont {Dahl}},
  \bibinfo {author} {\bibfnamefont {A.~M.}\ \bibnamefont {Hilger}}, \bibinfo
  {author} {\bibfnamefont {C.~J.}\ \bibnamefont {Ewasko}}, \ and\ \bibinfo
  {author} {\bibfnamefont {C.~M.}\ \bibnamefont {Hrenya}},\ }\href {\doibase
  10.1017/jfm.2013.328} {\bibfield  {journal} {\bibinfo  {journal} {J. Fluid
  Mech.}\ }\textbf {\bibinfo {volume} {729}},\ \unskip\ \bibinfo {pages}
  {484--495} (\bibinfo {year} {2013})}\BibitemShut {NoStop}%
\bibitem [{\citenamefont {{Vega Reyes}}, \citenamefont {Santos},\ and\
  \citenamefont {Kremer}(2014{\natexlab{a}})}]{VSK14}%
  \BibitemOpen
  \bibfield  {author} {\bibinfo {author} {\bibfnamefont {F.}~\bibnamefont
  {{Vega Reyes}}}, \bibinfo {author} {\bibfnamefont {A.}~\bibnamefont
  {Santos}}, \ and\ \bibinfo {author} {\bibfnamefont {G.~M.}\ \bibnamefont
  {Kremer}},\ }\href {\doibase 10.1103/PhysRevE.89.020202} {\bibfield
  {journal} {\bibinfo  {journal} {Phys. Rev. E}\ }\textbf {\bibinfo {volume}
  {89}},\ p.\ \bibinfo {pages} {020202(R)} (\bibinfo {year}
  {2014}{\natexlab{a}})}\BibitemShut {NoStop}%
\bibitem [{\citenamefont {{Vega Reyes}}, \citenamefont {Santos},\ and\
  \citenamefont {Kremer}(2014{\natexlab{b}})}]{VSK14b}%
  \BibitemOpen
  \bibfield  {author} {\bibinfo {author} {\bibfnamefont {F.}~\bibnamefont
  {{Vega Reyes}}}, \bibinfo {author} {\bibfnamefont {A.}~\bibnamefont
  {Santos}}, \ and\ \bibinfo {author} {\bibfnamefont {G.~M.}\ \bibnamefont
  {Kremer}},\ }\href {\doibase 10.1063/1.4902634} {\bibfield  {journal}
  {\bibinfo  {journal} {AIP Conf. Proc.}\ }\textbf {\bibinfo {volume} {1628}},\
  \unskip\ \bibinfo {pages} {494--501} (\bibinfo {year}
  {2014}{\natexlab{b}})}\BibitemShut {NoStop}%
\bibitem [{\citenamefont {Kremer}, \citenamefont {Santos},\ and\ \citenamefont
  {Garz\'o}(2014)}]{KSG14}%
  \BibitemOpen
  \bibfield  {author} {\bibinfo {author} {\bibfnamefont {G.~M.}\ \bibnamefont
  {Kremer}}, \bibinfo {author} {\bibfnamefont {A.}~\bibnamefont {Santos}}, \
  and\ \bibinfo {author} {\bibfnamefont {V.}~\bibnamefont {Garz\'o}},\ }\href
  {\doibase 10.1103/PhysRevE.90.022205} {\bibfield  {journal} {\bibinfo
  {journal} {Phys. Rev. E}\ }\textbf {\bibinfo {volume} {90}},\ p.\ \bibinfo
  {pages} {022205} (\bibinfo {year} {2014})}\BibitemShut {NoStop}%
\bibitem [{\citenamefont {Rongali}\ and\ \citenamefont {Alam}(2014)}]{RA14}%
  \BibitemOpen
  \bibfield  {author} {\bibinfo {author} {\bibfnamefont {R.}~\bibnamefont
  {Rongali}}\ and\ \bibinfo {author} {\bibfnamefont {M.}~\bibnamefont {Alam}},\
  }\href {\doibase 10.1103/PhysRevE.89.062201} {\bibfield  {journal} {\bibinfo
  {journal} {Phys. Rev. E}\ }\textbf {\bibinfo {volume} {89}},\ p.\ \bibinfo
  {pages} {062201} (\bibinfo {year} {2014})}\BibitemShut {NoStop}%
\bibitem [{\citenamefont {{Vega Reyes}}\ and\ \citenamefont
  {Santos}(2015)}]{VS15}%
  \BibitemOpen
  \bibfield  {author} {\bibinfo {author} {\bibfnamefont {F.}~\bibnamefont
  {{Vega Reyes}}}\ and\ \bibinfo {author} {\bibfnamefont {A.}~\bibnamefont
  {Santos}},\ }\href {\doibase 10.1063/1.4934727} {\bibfield  {journal}
  {\bibinfo  {journal} {Phys. Fluids}\ }\textbf {\bibinfo {volume} {27}},\ p.\
  \bibinfo {pages} {113301} (\bibinfo {year} {2015})}\BibitemShut {NoStop}%
\bibitem [{\citenamefont {Fullmer}\ and\ \citenamefont {Hrenya}(2017)}]{FH17}%
  \BibitemOpen
  \bibfield  {author} {\bibinfo {author} {\bibfnamefont {W.~D.}\ \bibnamefont
  {Fullmer}}\ and\ \bibinfo {author} {\bibfnamefont {C.~M.}\ \bibnamefont
  {Hrenya}},\ }\href {\doibase 10.1146/annurev-fluid-010816-060028} {\bibfield
  {journal} {\bibinfo  {journal} {Annu. Rev. Fluid Mech.}\ }\textbf {\bibinfo
  {volume} {49}},\ \unskip\ \bibinfo {pages} {485--510} (\bibinfo {year}
  {2017})}\BibitemShut {NoStop}%
\bibitem [{\citenamefont {Scholz}\ and\ \citenamefont
  {P\"oschel}(2017)}]{SP17}%
  \BibitemOpen
  \bibfield  {author} {\bibinfo {author} {\bibfnamefont {C.}~\bibnamefont
  {Scholz}}\ and\ \bibinfo {author} {\bibfnamefont {T.}~\bibnamefont
  {P\"oschel}},\ }\href {\doibase 10.1103/PhysRevLett.118.198003} {\bibfield
  {journal} {\bibinfo  {journal} {Phys. Rev. Lett.}\ }\textbf {\bibinfo
  {volume} {118}},\ p.\ \bibinfo {pages} {198003} (\bibinfo {year}
  {2017})}\BibitemShut {NoStop}%
\bibitem [{\citenamefont {Duan}\ and\ \citenamefont {Feng}(2017)}]{DF17}%
  \BibitemOpen
  \bibfield  {author} {\bibinfo {author} {\bibfnamefont {Y.}~\bibnamefont
  {Duan}}\ and\ \bibinfo {author} {\bibfnamefont {Z.-G.}\ \bibnamefont
  {Feng}},\ }\href {\doibase 10.1103/PhysRevE.96.062907} {\bibfield  {journal}
  {\bibinfo  {journal} {Phys. Rev. E}\ }\textbf {\bibinfo {volume} {96}},\ p.\
  \bibinfo {pages} {062907} (\bibinfo {year} {2017})}\BibitemShut {NoStop}%
\bibitem [{\citenamefont {Garz\'o}, \citenamefont {Santos},\ and\ \citenamefont
  {Kremer}(2018)}]{GSK18}%
  \BibitemOpen
  \bibfield  {author} {\bibinfo {author} {\bibfnamefont {V.}~\bibnamefont
  {Garz\'o}}, \bibinfo {author} {\bibfnamefont {A.}~\bibnamefont {Santos}}, \
  and\ \bibinfo {author} {\bibfnamefont {G.~M.}\ \bibnamefont {Kremer}},\
  }\href {\doibase 10.1103/PhysRevE.97.052901} {\bibfield  {journal} {\bibinfo
  {journal} {Phys. Rev. E}\ }\textbf {\bibinfo {volume} {97}},\ p.\ \bibinfo
  {pages} {052901} (\bibinfo {year} {2018})}\BibitemShut {NoStop}%
\bibitem [{\citenamefont {Jenkins}\ and\ \citenamefont {Mancini}(1989)}]{JM89}%
  \BibitemOpen
  \bibfield  {author} {\bibinfo {author} {\bibfnamefont {J.~T.}\ \bibnamefont
  {Jenkins}}\ and\ \bibinfo {author} {\bibfnamefont {F.}~\bibnamefont
  {Mancini}},\ }\href {\doibase 10.1063/1.857479} {\bibfield  {journal}
  {\bibinfo  {journal} {Phys. Fluids A}\ }\textbf {\bibinfo {volume} {1}},\
  \unskip\ \bibinfo {pages} {2050--2057} (\bibinfo {year} {1989})}\BibitemShut
  {NoStop}%
\bibitem [{\citenamefont {Garz\'o}\ and\ \citenamefont
  {Dufty}(1999{\natexlab{a}})}]{GD99b}%
  \BibitemOpen
  \bibfield  {author} {\bibinfo {author} {\bibfnamefont {V.}~\bibnamefont
  {Garz\'o}}\ and\ \bibinfo {author} {\bibfnamefont {J.~W.}\ \bibnamefont
  {Dufty}},\ }\href {\doibase 10.1103/PhysRevE.60.5706} {\bibfield  {journal}
  {\bibinfo  {journal} {Phys. Rev. E}\ }\textbf {\bibinfo {volume} {60}},\
  \unskip\ \bibinfo {pages} {5706--5713} (\bibinfo {year}
  {1999}{\natexlab{a}})}\BibitemShut {NoStop}%
\bibitem [{\citenamefont {Hong}, \citenamefont {Quinn},\ and\ \citenamefont
  {Luding}(2001)}]{HQL01}%
  \BibitemOpen
  \bibfield  {author} {\bibinfo {author} {\bibfnamefont {D.~C.}\ \bibnamefont
  {Hong}}, \bibinfo {author} {\bibfnamefont {P.~V.}\ \bibnamefont {Quinn}}, \
  and\ \bibinfo {author} {\bibfnamefont {S.}~\bibnamefont {Luding}},\ }\href
  {\doibase 10.1103/PhysRevLett.86.3423} {\bibfield  {journal} {\bibinfo
  {journal} {Phys. Rev. Lett.}\ }\textbf {\bibinfo {volume} {86}},\ \unskip\
  \bibinfo {pages} {3423--3426} (\bibinfo {year} {2001})}\BibitemShut {NoStop}%
\bibitem [{\citenamefont {Jenkins}\ and\ \citenamefont {Yoon}(2002)}]{JY02}%
  \BibitemOpen
  \bibfield  {author} {\bibinfo {author} {\bibfnamefont {J.~T.}\ \bibnamefont
  {Jenkins}}\ and\ \bibinfo {author} {\bibfnamefont {D.~K.}\ \bibnamefont
  {Yoon}},\ }\href {\doibase 10.1103/PhysRevLett.88.194301} {\bibfield
  {journal} {\bibinfo  {journal} {Phys. Rev. Lett.}\ }\textbf {\bibinfo
  {volume} {88}},\ p.\ \bibinfo {pages} {194301} (\bibinfo {year}
  {2002})}\BibitemShut {NoStop}%
\bibitem [{\citenamefont {Montanero}\ and\ \citenamefont
  {Garz\'o}(2002{\natexlab{a}})}]{MG02b}%
  \BibitemOpen
  \bibfield  {author} {\bibinfo {author} {\bibfnamefont {J.~M.}\ \bibnamefont
  {Montanero}}\ and\ \bibinfo {author} {\bibfnamefont {V.}~\bibnamefont
  {Garz\'o}},\ }\href {\doibase 10.1007/s10035-001-0097-8} {\bibfield
  {journal} {\bibinfo  {journal} {Granul. Matter}\ }\textbf {\bibinfo {volume}
  {4}},\ \unskip\ \bibinfo {pages} {17--24} (\bibinfo {year}
  {2002}{\natexlab{a}})}\BibitemShut {NoStop}%
\bibitem [{\citenamefont {Barrat}\ and\ \citenamefont {Trizac}(2002)}]{BT02a}%
  \BibitemOpen
  \bibfield  {author} {\bibinfo {author} {\bibfnamefont {A.}~\bibnamefont
  {Barrat}}\ and\ \bibinfo {author} {\bibfnamefont {E.}~\bibnamefont
  {Trizac}},\ }\href {\doibase 10.1007/s10035-002-0108-4} {\bibfield  {journal}
  {\bibinfo  {journal} {Granul. Matter}\ }\textbf {\bibinfo {volume} {4}},\
  \unskip\ \bibinfo {pages} {57--63} (\bibinfo {year} {2002})}\BibitemShut
  {NoStop}%
\bibitem [{\citenamefont {Dahl}\ \emph {et~al.}(2002)\citenamefont {Dahl},
  \citenamefont {Hrenya}, \citenamefont {Garz\'o},\ and\ \citenamefont
  {Dufty}}]{DHGD02}%
  \BibitemOpen
  \bibfield  {author} {\bibinfo {author} {\bibfnamefont {S.~R.}\ \bibnamefont
  {Dahl}}, \bibinfo {author} {\bibfnamefont {C.~M.}\ \bibnamefont {Hrenya}},
  \bibinfo {author} {\bibfnamefont {V.}~\bibnamefont {Garz\'o}}, \ and\
  \bibinfo {author} {\bibfnamefont {J.~W.}\ \bibnamefont {Dufty}},\ }\href
  {\doibase 10.1103/PhysRevE.66.041301} {\bibfield  {journal} {\bibinfo
  {journal} {Phys. Rev. E}\ }\textbf {\bibinfo {volume} {66}},\ p.\ \bibinfo
  {pages} {041301} (\bibinfo {year} {2002})}\BibitemShut {NoStop}%
\bibitem [{\citenamefont {Garz\'o}\ and\ \citenamefont {Dufty}(2002)}]{GD02}%
  \BibitemOpen
  \bibfield  {author} {\bibinfo {author} {\bibfnamefont {V.}~\bibnamefont
  {Garz\'o}}\ and\ \bibinfo {author} {\bibfnamefont {J.~W.}\ \bibnamefont
  {Dufty}},\ }\href {\doibase 10.1063/1.1458007} {\bibfield  {journal}
  {\bibinfo  {journal} {Phys. Fluids}\ }\textbf {\bibinfo {volume} {14}},\
  \unskip\ \bibinfo {pages} {1476--1490} (\bibinfo {year} {2002})}\BibitemShut
  {NoStop}%
\bibitem [{\citenamefont {Garz\'o}(2002)}]{G02}%
  \BibitemOpen
  \bibfield  {author} {\bibinfo {author} {\bibfnamefont {V.}~\bibnamefont
  {Garz\'o}},\ }\href {\doibase 10.1103/PhysRevE.66.021308} {\bibfield
  {journal} {\bibinfo  {journal} {Phys. Rev. E}\ }\textbf {\bibinfo {volume}
  {66}},\ p.\ \bibinfo {pages} {021308} (\bibinfo {year} {2002})}\BibitemShut
  {NoStop}%
\bibitem [{\citenamefont {Brey}, \citenamefont {Ruiz-Montero},\ and\
  \citenamefont {Moreno}(2005)}]{BRM05}%
  \BibitemOpen
  \bibfield  {author} {\bibinfo {author} {\bibfnamefont {J.~J.}\ \bibnamefont
  {Brey}}, \bibinfo {author} {\bibfnamefont {M.~J.}\ \bibnamefont
  {Ruiz-Montero}}, \ and\ \bibinfo {author} {\bibfnamefont {F.}~\bibnamefont
  {Moreno}},\ }\href {\doibase http://dx.doi.org/10.1103/PhysRevLett.95.098001}
  {\bibfield  {journal} {\bibinfo  {journal} {Phys. Rev. Lett.}\ }\textbf
  {\bibinfo {volume} {95}},\ p.\ \bibinfo {pages} {098001} (\bibinfo {year}
  {2005})}\BibitemShut {NoStop}%
\bibitem [{\citenamefont {Santos}\ and\ \citenamefont
  {Astillero}(2005)}]{SA05}%
  \BibitemOpen
  \bibfield  {author} {\bibinfo {author} {\bibfnamefont {A.}~\bibnamefont
  {Santos}}\ and\ \bibinfo {author} {\bibfnamefont {A.}~\bibnamefont
  {Astillero}},\ }\href {\doibase 10.1103/PhysRevE.72.031308} {\bibfield
  {journal} {\bibinfo  {journal} {Phys. Rev. E}\ }\textbf {\bibinfo {volume}
  {72}},\ p.\ \bibinfo {pages} {031308} (\bibinfo {year} {2005})}\BibitemShut
  {NoStop}%
\bibitem [{\citenamefont {Serero}\ \emph {et~al.}(2006)\citenamefont {Serero},
  \citenamefont {Goldhirsch}, \citenamefont {Noskowicz},\ and\ \citenamefont
  {Tan}}]{SGNT06}%
  \BibitemOpen
  \bibfield  {author} {\bibinfo {author} {\bibfnamefont {D.}~\bibnamefont
  {Serero}}, \bibinfo {author} {\bibfnamefont {I.}~\bibnamefont {Goldhirsch}},
  \bibinfo {author} {\bibfnamefont {S.~H.}\ \bibnamefont {Noskowicz}}, \ and\
  \bibinfo {author} {\bibfnamefont {M.-L.}\ \bibnamefont {Tan}},\ }\href
  {\doibase 10.1017/S0022112006009281} {\bibfield  {journal} {\bibinfo
  {journal} {J. Fluid Mech.}\ }\textbf {\bibinfo {volume} {554}},\ \unskip\
  \bibinfo {pages} {237--258} (\bibinfo {year} {2006})}\BibitemShut {NoStop}%
\bibitem [{\citenamefont {Garz\'o}, \citenamefont {Dufty},\ and\ \citenamefont
  {Hrenya}(2007)}]{GDH07}%
  \BibitemOpen
  \bibfield  {author} {\bibinfo {author} {\bibfnamefont {V.}~\bibnamefont
  {Garz\'o}}, \bibinfo {author} {\bibfnamefont {J.~W.}\ \bibnamefont {Dufty}},
  \ and\ \bibinfo {author} {\bibfnamefont {C.~M.}\ \bibnamefont {Hrenya}},\
  }\href {\doibase 10.1103/PhysRevE.76.031303} {\bibfield  {journal} {\bibinfo
  {journal} {Phys. Rev. E}\ }\textbf {\bibinfo {volume} {76}},\ p.\ \bibinfo
  {pages} {031303} (\bibinfo {year} {2007})}\BibitemShut {NoStop}%
\bibitem [{\citenamefont {Garz\'o}, \citenamefont {Hrenya},\ and\ \citenamefont
  {Dufty}(2007)}]{GHD07}%
  \BibitemOpen
  \bibfield  {author} {\bibinfo {author} {\bibfnamefont {V.}~\bibnamefont
  {Garz\'o}}, \bibinfo {author} {\bibfnamefont {C.~M.}\ \bibnamefont {Hrenya}},
  \ and\ \bibinfo {author} {\bibfnamefont {J.~W.}\ \bibnamefont {Dufty}},\
  }\href {\doibase 10.1103/PhysRevE.76.031304} {\bibfield  {journal} {\bibinfo
  {journal} {Phys. Rev. E}\ }\textbf {\bibinfo {volume} {76}},\ p.\ \bibinfo
  {pages} {031304} (\bibinfo {year} {2007})}\BibitemShut {NoStop}%
\bibitem [{\citenamefont {Garz\'o}\ and\ \citenamefont
  {Montanero}(2007)}]{GM07}%
  \BibitemOpen
  \bibfield  {author} {\bibinfo {author} {\bibfnamefont {V.}~\bibnamefont
  {Garz\'o}}\ and\ \bibinfo {author} {\bibfnamefont {J.~M.}\ \bibnamefont
  {Montanero}},\ }\href {\doibase 10.1007/s10955-007-9357-2} {\bibfield
  {journal} {\bibinfo  {journal} {J. Stat. Phys.}\ }\textbf {\bibinfo {volume}
  {129}},\ \unskip\ \bibinfo {pages} {27--58} (\bibinfo {year}
  {2007})}\BibitemShut {NoStop}%
\bibitem [{\citenamefont {{Vega Reyes}}, \citenamefont {Garz\'o},\ and\
  \citenamefont {Santos}(2007)}]{VGS07}%
  \BibitemOpen
  \bibfield  {author} {\bibinfo {author} {\bibfnamefont {F.}~\bibnamefont
  {{Vega Reyes}}}, \bibinfo {author} {\bibfnamefont {V.}~\bibnamefont
  {Garz\'o}}, \ and\ \bibinfo {author} {\bibfnamefont {A.}~\bibnamefont
  {Santos}},\ }\href {\doibase 10.1103/PhysRevE.75.061306} {\bibfield
  {journal} {\bibinfo  {journal} {Phys. Rev. E}\ }\textbf {\bibinfo {volume}
  {75}},\ p.\ \bibinfo {pages} {061306} (\bibinfo {year} {2007})}\BibitemShut
  {NoStop}%
\bibitem [{\citenamefont {Garz\'o}(2008{\natexlab{a}})}]{G08}%
  \BibitemOpen
  \bibfield  {author} {\bibinfo {author} {\bibfnamefont {V.}~\bibnamefont
  {Garz\'o}},\ }\href {\doibase 10.1103/PhysRevE.78.020301} {\bibfield
  {journal} {\bibinfo  {journal} {Phys. Rev. E}\ }\textbf {\bibinfo {volume}
  {78}},\ p.\ \bibinfo {pages} {020301(R)} (\bibinfo {year}
  {2008}{\natexlab{a}})}\BibitemShut {NoStop}%
\bibitem [{\citenamefont {Garz\'o}(2008{\natexlab{b}})}]{G08b}%
  \BibitemOpen
  \bibfield  {author} {\bibinfo {author} {\bibfnamefont {V.}~\bibnamefont
  {Garz\'o}},\ }\enquote {\bibinfo {title} {{Kinetic Theory for Binary Granular
  Mixtures at Low Density}},}\ in\ \href@noop {} {\emph {\bibinfo {booktitle}
  {{Theory and Simulation of Hard-Sphere Fluids and Related Systems}}}},\
  \bibinfo {series} {Lectures Notes in Physics}, Vol.\ \bibinfo {volume}
  {753},\ \bibinfo {editor} {edited by\ \bibinfo {editor} {\bibfnamefont
  {A.}~\bibnamefont {Mulero}}}\ (\bibinfo  {publisher} {Springer-Verlag},\
  \bibinfo {address} {Berlin},\ \bibinfo {year} {2008})\ \unskip, pp.\ \bibinfo
  {pages} {493--540}\BibitemShut {NoStop}%
\bibitem [{\citenamefont {Uecker}\ \emph {et~al.}(2009)\citenamefont {Uecker},
  \citenamefont {Kranz}, \citenamefont {Aspelmeier},\ and\ \citenamefont
  {Zippelius}}]{UKAZ09}%
  \BibitemOpen
  \bibfield  {author} {\bibinfo {author} {\bibfnamefont {H.}~\bibnamefont
  {Uecker}}, \bibinfo {author} {\bibfnamefont {W.~T.}\ \bibnamefont {Kranz}},
  \bibinfo {author} {\bibfnamefont {T.}~\bibnamefont {Aspelmeier}}, \ and\
  \bibinfo {author} {\bibfnamefont {A.}~\bibnamefont {Zippelius}},\ }\href
  {\doibase 10.1103/PhysRevE.80.041303} {\bibfield  {journal} {\bibinfo
  {journal} {Phys. Rev. E}\ }\textbf {\bibinfo {volume} {80}},\ p.\ \bibinfo
  {pages} {041303} (\bibinfo {year} {2009})}\BibitemShut {NoStop}%
\bibitem [{\citenamefont {Viot}\ and\ \citenamefont {Talbot}(2004)}]{VT04}%
  \BibitemOpen
  \bibfield  {author} {\bibinfo {author} {\bibfnamefont {P.}~\bibnamefont
  {Viot}}\ and\ \bibinfo {author} {\bibfnamefont {J.}~\bibnamefont {Talbot}},\
  }\href {\doibase 10.1103/PhysRevE.69.051106} {\bibfield  {journal} {\bibinfo
  {journal} {Phys. Rev. E}\ }\textbf {\bibinfo {volume} {69}},\ p.\ \bibinfo
  {pages} {051106} (\bibinfo {year} {2004})}\BibitemShut {NoStop}%
\bibitem [{\citenamefont {Piasecki}, \citenamefont {Talbot},\ and\
  \citenamefont {Viot}(2007{\natexlab{a}})}]{PTV07}%
  \BibitemOpen
  \bibfield  {author} {\bibinfo {author} {\bibfnamefont {J.}~\bibnamefont
  {Piasecki}}, \bibinfo {author} {\bibfnamefont {J.}~\bibnamefont {Talbot}}, \
  and\ \bibinfo {author} {\bibfnamefont {P.}~\bibnamefont {Viot}},\ }\href
  {\doibase 10.1103/PhysRevE.75.051307} {\bibfield  {journal} {\bibinfo
  {journal} {Phys. Rev. E}\ }\textbf {\bibinfo {volume} {75}},\ p.\ \bibinfo
  {pages} {051307} (\bibinfo {year} {2007}{\natexlab{a}})}\BibitemShut
  {NoStop}%
\bibitem [{\citenamefont {Cornu}\ and\ \citenamefont {Piasecki}(2008)}]{CP08}%
  \BibitemOpen
  \bibfield  {author} {\bibinfo {author} {\bibfnamefont {F.}~\bibnamefont
  {Cornu}}\ and\ \bibinfo {author} {\bibfnamefont {J.}~\bibnamefont
  {Piasecki}},\ }\href {\doibase 10.1016/j.physa.2008.03.014} {\bibfield
  {journal} {\bibinfo  {journal} {Physica A}\ }\textbf {\bibinfo {volume}
  {387}},\ \unskip\ \bibinfo {pages} {4856--4862} (\bibinfo {year}
  {2008})}\BibitemShut {NoStop}%
\bibitem [{\citenamefont {Santos}, \citenamefont {Kremer},\ and\ \citenamefont
  {Garz\'o}(2010)}]{SKG10}%
  \BibitemOpen
  \bibfield  {author} {\bibinfo {author} {\bibfnamefont {A.}~\bibnamefont
  {Santos}}, \bibinfo {author} {\bibfnamefont {G.~M.}\ \bibnamefont {Kremer}},
  \ and\ \bibinfo {author} {\bibfnamefont {V.}~\bibnamefont {Garz\'o}},\ }\href
  {\doibase 10.1143/PTPS.184.31} {\bibfield  {journal} {\bibinfo  {journal}
  {Prog. Theor. Phys. Suppl.}\ }\textbf {\bibinfo {volume} {184}},\ \unskip\
  \bibinfo {pages} {31--48} (\bibinfo {year} {2010})}\BibitemShut {NoStop}%
\bibitem [{\citenamefont {Santos}(2011{\natexlab{b}})}]{S11b}%
  \BibitemOpen
  \bibfield  {author} {\bibinfo {author} {\bibfnamefont {A.}~\bibnamefont
  {Santos}},\ }\href {\doibase 10.1063/1.3562637} {\bibfield  {journal}
  {\bibinfo  {journal} {AIP Conf. Proc.}\ }\textbf {\bibinfo {volume} {1333}},\
  \unskip\ \bibinfo {pages} {128--133} (\bibinfo {year}
  {2011}{\natexlab{b}})}\BibitemShut {NoStop}%
\bibitem [{\citenamefont {{Vega Reyes}}\ \emph
  {et~al.}(2017{\natexlab{a}})\citenamefont {{Vega Reyes}}, \citenamefont
  {Lasanta}, \citenamefont {Santos},\ and\ \citenamefont {Garz\'o}}]{VLSG17}%
  \BibitemOpen
  \bibfield  {author} {\bibinfo {author} {\bibfnamefont {F.}~\bibnamefont
  {{Vega Reyes}}}, \bibinfo {author} {\bibfnamefont {A.}~\bibnamefont
  {Lasanta}}, \bibinfo {author} {\bibfnamefont {A.}~\bibnamefont {Santos}}, \
  and\ \bibinfo {author} {\bibfnamefont {V.}~\bibnamefont {Garz\'o}},\ }\href
  {\doibase 10.1051/epjconf/201714004003} {\bibfield  {journal} {\bibinfo
  {journal} {EPJ Web Conf.}\ }\textbf {\bibinfo {volume} {140}},\ p.\ \bibinfo
  {pages} {04003} (\bibinfo {year} {2017}{\natexlab{a}})}\BibitemShut {NoStop}%
\bibitem [{\citenamefont {{Vega Reyes}}\ \emph
  {et~al.}(2017{\natexlab{b}})\citenamefont {{Vega Reyes}}, \citenamefont
  {Lasanta}, \citenamefont {Santos},\ and\ \citenamefont {Garz\'o}}]{VLSG17b}%
  \BibitemOpen
  \bibfield  {author} {\bibinfo {author} {\bibfnamefont {F.}~\bibnamefont
  {{Vega Reyes}}}, \bibinfo {author} {\bibfnamefont {A.}~\bibnamefont
  {Lasanta}}, \bibinfo {author} {\bibfnamefont {A.}~\bibnamefont {Santos}}, \
  and\ \bibinfo {author} {\bibfnamefont {V.}~\bibnamefont {Garz\'o}},\ }\href
  {\doibase 10.1103/PhysRevE.96.052901} {\bibfield  {journal} {\bibinfo
  {journal} {Phys. Rev. E}\ }\textbf {\bibinfo {volume} {96}},\ p.\ \bibinfo
  {pages} {052901} (\bibinfo {year} {2017}{\natexlab{b}})}\BibitemShut
  {NoStop}%
\bibitem [{\citenamefont {Santos}(2018)}]{S18}%
  \BibitemOpen
  \bibfield  {author} {\bibinfo {author} {\bibfnamefont {A.}~\bibnamefont
  {Santos}},\ }\href {\doibase 10.1103/PhysRevE.98.012904} {\bibfield
  {journal} {\bibinfo  {journal} {Phys. Rev. E}\ }\textbf {\bibinfo {volume}
  {98}},\ p.\ \bibinfo {pages} {012804} (\bibinfo {year} {2018})}\BibitemShut
  {NoStop}%
\bibitem [{\citenamefont {Brey}\ and\ \citenamefont {Cubero}(2001)}]{BC01}%
  \BibitemOpen
  \bibfield  {author} {\bibinfo {author} {\bibfnamefont {J.~J.}\ \bibnamefont
  {Brey}}\ and\ \bibinfo {author} {\bibfnamefont {D.}~\bibnamefont {Cubero}},\
  }\enquote {\bibinfo {title} {Hydrodynamic transport coefficients of granular
  gases},}\ in\ \href@noop {} {\emph {\bibinfo {booktitle} {Granular Gases}}},\
  \bibinfo {series} {Lectures Notes in Physics}, Vol.\ \bibinfo {volume}
  {564},\ \bibinfo {editor} {edited by\ \bibinfo {editor} {\bibfnamefont
  {T.}~\bibnamefont {P\"oschel}}\ and\ \bibinfo {editor} {\bibfnamefont
  {S.}~\bibnamefont {Luding}}}\ (\bibinfo  {publisher} {Springer},\ \bibinfo
  {address} {Berlin},\ \bibinfo {year} {2001})\ \unskip, pp.\ \bibinfo {pages}
  {59--78}\BibitemShut {NoStop}%
\bibitem [{\citenamefont {van Noije}\ and\ \citenamefont
  {Ernst}(1998)}]{vNE98}%
  \BibitemOpen
  \bibfield  {author} {\bibinfo {author} {\bibfnamefont {T.~P.~C.}\
  \bibnamefont {van Noije}}\ and\ \bibinfo {author} {\bibfnamefont {M.~H.}\
  \bibnamefont {Ernst}},\ }\href@noop {} {\bibfield  {journal} {\bibinfo
  {journal} {Granul. Matter}\ }\textbf {\bibinfo {volume} {1}},\ \unskip\
  \bibinfo {pages} {57--64} (\bibinfo {year} {1998})}\BibitemShut {NoStop}%
\bibitem [{\citenamefont {Garz{\'o}}\ and\ \citenamefont
  {Santos}(2003)}]{GS03}%
  \BibitemOpen
  \bibfield  {author} {\bibinfo {author} {\bibfnamefont {V.}~\bibnamefont
  {Garz{\'o}}}\ and\ \bibinfo {author} {\bibfnamefont {A.}~\bibnamefont
  {Santos}},\ }\href@noop {} {\emph {\bibinfo {title} {Kinetic Theory of Gases
  in Shear Flows: Nonlinear Transport}}},\ Fundamental Theories of Physics\
  (\bibinfo  {publisher} {Springer},\ \bibinfo
  {address} {Dordrecht},\ \bibinfo {year}
  {2003})\BibitemShut {NoStop}%
\end{thebibliography}

%Control: year  (1) truncated

\end{document}